
\headline={\ifnum\pageno=1\firstheadline\else
\ifodd\pageno\rightheadline \else\leftheadline\fi\fi}
\def\firstheadline{\hfil}
\def\rightheadline{\hfil}
\def\leftheadline{\hfil}
	\footline={\ifnum\pageno=1\firstfootline\else\otherfootline\fi}
\def\firstfootline{\rm\hss\folio\hss}
\def\otherfootline{\hfil}

\font\twelvebf=cmbx10 scaled\magstep 1
\font\twelverm=cmr10 scaled\magstep 1
\font\twelveit=cmti10 scaled\magstep 1

\font\tenbf=cmbx10
\font\tenrm=cmr10
\font\tenit=cmti10

\font\ninerm=cmr9

\parindent=1.5pc
\hsize=6.0truein
\vsize=8.5truein
\nopagenumbers
\line{\hfil \ninerm  CCNY-HEP-94/10}
\line{\hfil \ninerm  October 1994}
\medskip
\centerline{\tenbf CHERN-SIMONS AND WZNW THEORIES AND}
\baselineskip=22pt
\centerline{\bf THE QUARK-GLUON PLASMA {\ninerm \footnote{*} {
Lectures presented at Mt. Sorak Summer School, S. Korea, June 27-July 2,
1994}}}
\baselineskip=16pt
\vglue 0.8cm
\centerline{\tenrm V.P. NAIR }
\baselineskip=13pt
\centerline{\tenit Physics Department }
\centerline{\tenit City College of the City University of New York}
\baselineskip=12pt
\centerline{\tenit New York, New York 10031, U.S.A.}
\vglue 0.8cm
\centerline{\tenrm ABSTRACT}
\vglue 0.3cm
{\rightskip=3pc
 \leftskip=3pc
 \tenrm\baselineskip=12pt\noindent
Summation over hard thermal loops, by themselves and as
insertions in higher order Feynman diagrams, is important in
thermal perturbation theory for Quantum Chromodynamics, so that
all contributions of a given order in the coupling constant can be
consistently taken into account. I review some of the basic
properties of hard thermal loops and how the generating functional
for them is related to the eikonal for a
Chern-Simons gauge
theory, and using an auxiliary field, to the gauged
Wess-Zumino-Novikov-Witten action. The Hamiltonian analysis of
the effective action and a discussion of plasma waves are given.
It is also pointed out that a possible expression for the
magnetic mass term can be written in a closely related way.
\vglue 0.6cm}
\vfil
\twelverm
\baselineskip=14pt
\vglue 0.3cm
\centerline{\twelvebf Contents}
\vskip .2in
\leftline{\twelveit 1. Introduction}
\vskip .1in
\noindent{\twelveit 2. Analysis of diagrams and power counting rules}
\vskip .1in
\noindent{\twelveit 3. Calculation of the two-point function}
\vskip .1in
\noindent{\twelveit 4. Properties of $\Gamma [A]$}
\vskip .1in
\noindent{\twelveit 5. Chern-Simons and WZNW theories}
\vskip .1in
\noindent{\twelveit 6. Solution for $\Gamma [A]$}
\vskip .1in
\noindent{\twelveit 7. Continuation to Minkowski space }
\vskip .1in
\noindent{\twelveit 8. Derivation of $\Gamma [A]$ from the Boltzmann equation}
\vskip .1in
\noindent{\twelveit 9. An auxiliary field and  Hamitonian analysis}
\vskip .1in
\noindent{\twelveit 10. Plasma waves}
\vskip .1in
\noindent{\twelveit 11. Magnetic screening and magnetic mass}
\vfil\eject

\def\tr{{\rm tr}}
\def\Tr{{\rm Tr}}
\def\G {{\Gamma}}
\def\gA{{\gamma \cdot A}}
\def\gp{{\gamma \cdot p}}
\def\gq{{\gamma \cdot q}}

\def\ap{{\alpha_p}}
\def\aq{{\alpha_q}}

\def\bp{{\beta_p}}
\def\bq{{\beta_q}}

\def\abp{{\bar \alpha_p}}
\def\abq{{\bar \alpha_q}}

\def\bbp{{\bar \beta_p}}
\def\bbq{{\bar \beta_q}}

\def\iep {{i\epsilon}}
\def\aoq{A_1 \cdot Q}
\def\atq{A_2 \cdot Q}

\def\kq{{k \cdot Q}}

\def\vqk{{\vec Q \cdot \vec k}}
\def\dq{{\cdot Q}}

\def\vq{{\vec q}}
\def\vk {{\vec k}}
\def\vQ{{\vec Q}}
\def\vp {{\vec p}}
\def\vx {{\vec x}}
\def\vy {{\vec y}}
\def \k {{\kappa}}
\def\12 {{\textstyle {1\over 2}}}
\noindent{\twelvebf 1. Introduction}
\vskip .2in
In these lectures, I review the relationship
between Chern-Simons (CS) and Wess-Zumino-Novikov-Witten (WZNW)
theories and Quantum Chromodynamics (QCD) at finite
temperature $^{1-3}$.
The results are expected to be useful in understanding the theory
of the quark-gluon plasma.

There is essentially universal agreement by now that
QCD is the theory
of strong interactions. One of the interesting features of QCD is
the deconfinement phase transition which is indicated by general
theoretical arguments as well as lattice simulations $^{4,5}$.
If the ambient temperature is raised beyond a critical value $T_C$,
hadronic matter makes a phase transition. Quarks and gluons are no
longer confined; we have a plasma of quarks, antiquarks and gluons.
As a new phase of hadronic matter, the quark-gluon plasma is of
considerable interest to physicists.

The plasma phase of hadronic matter, it is generally believed,
can be achieved in the collisions of
sufficiently heavy nuclei at sufficiently high energy. However,
the key word here
is {\it sufficiently}. Will the relativistic heavy ion collider
(RHIC), to be
built at Brookhaven, have enough energy to achieve this? Can thermal
equilibrium be obtained even for a short duration before the plasma
cools and rehadronizes? These are questions to which, as of now,
there are no definitive and clear answers $^6$.
Nevertheless, understanding physical
phenomena in the quark-gluon plasma at and near thermal equilibrium,
which can be described by QCD at high temperatures, is an important
and necessary first step. The critical temperature $T_C$
for the transition
to the plasma phase is expected to be of the order of the
scale parameter, $\Lambda_{QCD}$ which is around 200 MeV. Physical
phenomena near this temperature are of great interest, especially
since they may serve as signatures for the transition $^7$.
We shall
not discuss these however. Instead, we shall consider temperatures
which are significantly higher than $T_C$; the average momentum
carried by quarks and gluons will be much higher than
$\Lambda_{QCD}$ and one can expect thermal perturbation theory
to provide an adequate description of the physics. We are
interested in the technical problems of thermal perturbation
theory for QCD, especially hard thermal loops and related
physical phenomena such as Debye screening, Landau damping,
propagation of plasma waves, etc. (Eventhough we consider
$T \gg T_C$,
many of the results are of importance near $T_C$ as well. It is
qualitatively easier to isolate the relevant terms
 at high temperatures;
the modifications required at lower temperatures can be easily
included. The quark flavors of interest are also the light quarks,
with masses small compared to $T$. The heavier flavors have the
standard Boltzmann suppression factors and lead to small
modifications.)

As is well known , among the most elementary phenomena in a plasma
are Debye screening and Landau damping.
If we consider an Abelian plasma of
positive and negative charges $\pm e$, viz., electrodynamics,
screening can be understood using the classic argument of Debye.
One considers the Poisson equation for the electrostatic potential of a
test charge, say positive, in the plasma.
$$
\eqalignno{
-\nabla^2~A_0~&=~ ne\left( {{e^{eA_0 /T}~-~e^{-eA_0/T}}\over{e^{eA_0/T}~+~
e^{-eA_0/T}}}\right) &(1a)\cr
&\approx \left( ne^2\over T\right)A_0 &(1b) \cr}
$$
where the right hand side is the charge density in the vicinity of the
test charge. $n$ is the average number density of particles. The
exponentials are the Boltzmann factors giving the preferential
accumulation of negative and depletion of positive charges in the
vicinity due to the Coulomb forces, with proper normalization.
The approximation (1b), which is valid for high temperatures, shows that
the solutions have the screened Coulomb form $(1/r)\exp(-m_D r)$ with
a Debye screening mass $m_D^2=(ne^2/T)$. For a relativistic plasma,
the qualitative features of this argument are valid and with $n\sim
T^3$, we expect $m_D^2\sim e^2 T^2$. And by calculating the photon
propagator in thermal electrodynamics, one can indeed obtain a
similar result $^8$.

The above argument is simple and nice, but is
presented in terms of potentials and as such, it
does not seem to be gauge invariant. For the Abelian case, one can
easily reformulate the arguments using only the gauge-invariant electric
and magnetic fields. However, in a non-Abelian plasma, such as the
quark-gluon plasma, it is difficult to avoid the use of gauge potentials
altogether. Further, since even the notion of the charge of a gluon has
to be defined with respect to some chosen Abelian direction of the gauge
group, it is clear that the simple argument of Eq.(1) will have to
be modified. The question of interest to us is: how do we obtain a
gauge-invariant description of Debye screening in a non-Abelian
plasma (in terms of gauge potentials)?
More specifically, we need a functional of the
gauge potentials, $\Gamma [A]$, which is generated by the statistical
distributions and is effectively a gauge-invariant mass term for the
gauge fields.
Screening, of course, is the static part of a more general problem, the
dynamical part of which is the propagation of plasma waves. Such a $\Gamma [A]$
will thus be
important for the discussion of plasma waves as well.

Landau damping arises from the scattering of the particles in the
plasma against
an external field. This can transfer energy and momentum
from the field to the particle, leading to damping of the field.
The process is essentially the inverse of
${\rm{\check C}erenkov}$ radiation and can be
described by the proper continuation of the results for plasma waves
to a region of spacelike momenta.
Of course, being a time-asymmetric process, one must describe it
in terms of the equations of motion rather than an action; the
relevant current for these
equations will be a continuation of what is given by $\Gamma [A]$.

Hard thermal loops are closely related to the above. They arise because
of the need to carry out a partial resummation of perturbation theory in
thermal QCD. The need for resummation is easily seen using the following
argument due to Pisarski $^9$. Consider the elementary polarization diagram
of gluons, fig.(1a) and its first correction fig.(1b). Before the final
loop integration over $p$, the ratio of diagram (1b) to (1a) is $\Pi
(p)/p^2$. Because of Debye screening, the polarization tensor $\Pi (p)$
has a small $p$-expansion, $\Pi (p) \sim g^2 T^2 ~+~ g^2 T \vert
p\vert~+...$. ($g$ is the quark-gluon or gluon-gluon coupling constant.)
We thus see that for the small $p$-regime of integration, i.e.,
 $p^2 \leq g^2T^2$, the naively higher order diagram (1b) is
comparable to the
lower order term (1a). Therefore, to be consistent to a given order
in $g$, one must sum over (1a), (1b), and a series of further
insertions of
$\Pi (p)$. This resummation leads to new effective propagators;
in general,
we shall need new effective vertices as well.
The resummation can be summarized by
saying that the new propagators and vertices arise from an action
$$
S~=~ \int d^4x~ \left({\textstyle -{1\over 4}}F^2 \right)
{}~+~\Gamma [A]\eqno(2)
$$
where we add a term $\Gamma [A]$ to the standard Yang-Mills action
for gluons.

Now, going back to figures (1), we see that the regime of interest is
when the momentum external to $\Pi (p)$, viz., $p$ is of the order
$gT$ or less. We can thus take the momenta carried by
the potentials in
$\Gamma [A]$  to be ${\buildrel < \over \sim} gT$.
In order to determine which diagrams can contribute to
$\Gamma [A]$, we shall need to extend the
argument given above for $\Pi (p)$ to general vertices and obtain
rules for counting powers of $g$ and $T$,
keeping in mind the
restriction on external momenta. These rules basically follow from
the thermal propagator and let us briefly examine them.
\vskip .6cm
\noindent {\twelvebf 2. Analysis of diagrams and power counting rules}
\vskip .2in
The propagator for a quark at finite temperature is defined by
$$
S(x,y) = \langle {\cal T }~q(x) \bar q(y) \rangle  \eqno(3)
$$
where ${\cal T}$ denotes time-ordering. We shall use Minkowski
space propagators with thermal averages for the products of creation
and annihilation operators;
this is conceptually the simplest for
us since we need to talk about nonequilibrium phenomena such as
the propagation of plasma waves.
The propagator can be evaluated as
$$\eqalign{ S(x,y) = \int {d^3p \over (2\pi )^3}{1 \over 2p^0} \lbrace
            &\Theta (x^0 - y^0) \lbrack \ap e^{-i p \cdot (x-y)} \gp
            +\bbp e^{i p' \cdot (x-y)}  \gp' \rbrack - \cr
            & \Theta (y^0- x^0) \lbrack \bp e^{-i p \cdot (x-y)} \gp
            +\abp e^{i p' \cdot (x-y)}  \gp' \rbrack \rbrace \cr}
 \eqno(4)
$$

\noindent where $p^0 = |\vec p|, \; p = (p^0,\vec p),
\; p' = (p^0,-\vec p),
$
and $\Theta(x)$, of course, is the step function. Also
$$ \ap = 1 - n_p,  ~~~~~~\bp = n_p.  \eqno(5)$$

The distribution functions $n_p, \bar n_p$ are defined by the thermal
averages
$$\eqalignno {\langle  a_p^{\dag \alpha,r} a_p^{\beta,s} \rangle& = n_p
   \delta^{rs} \delta^{\alpha \beta} \cr
   \langle  b_p^{\dag \alpha,r} b_p^{\beta,s} \rangle &=\bar n_p
   \delta^{rs} \delta^{\alpha \beta} &(6) \cr} $$

\noindent where $(a_p^{\dag \alpha,r} , a_p^{\alpha,r}),
        (b_p^{\dag \alpha,r} , b_p^{\alpha,r})$ are the annihilation and
creation operators for quarks and antiquarks respectively.
$\alpha,\beta$ are spin indices; $r,s$ are color indices.
For a plasma of zero fermion number, we can take
$$ n_p = \bar n_p = {1 \over e^{p^0/T} + 1}  \eqno(7) $$
For a plasma with a nonzero value of fermion number, there is a
chemical
potential and correspondingly $n_p, \bar n_p$ are not equal.

We can also write the propagator, for zero chemical potential, as
$$
S(x,y) = ~\int {{d^4p}\over {(2\pi)^4}}~e^{-ip(x-y)} \left[ i{\gp
\over {p^2+i\epsilon}} ~-~ \gp ~2\pi ~n_p \delta (p^2 ) \right]
\eqno(8)
$$
The first term in brackets is the standard Feynman term at $T=0$.
The $T$-dependent correction is effectively on mass-shell, enforced by
the $\delta$-function.

For the gluon field, the propagator in the Feynman gauge
can be similarly written as
$G^{ab}_{\mu\nu} (x,y)= \delta^{ab} g_{\mu\nu} G(x,y)$ where
$g_{\mu\nu}$ is the metric tensor, $a,b$ are color indices and
$$
\eqalignno{
G(x,y)~&= ~ \int {d^3p \over (2\pi )^3}{1 \over 2p^0}
\Biggl[[ \Theta (x^0-y^0)~\{ \alpha_p e^{-ip(x-y)}  +
\beta_p e^{ip(x-y)}\}&{}\cr
&~~~~~~+ \Theta (y^0-x^0)~\{ \beta_p e^{-ip(x-y)} +
\alpha_p e^{ip(x-y)}\}\Biggr]&(9a)\cr
&= \int {{d^4p }\over {(2\pi)^4}} \left[ {i\over {p^2+\iep}} ~+~
2\pi ~n_p ~\delta (p^2)\right] &(9b)\cr}
$$
Here $\alpha_p= 1+n_p, ~\beta_p= n_p$ and $n_p$ is the
bosonic  distribution function
$$
n_p = {1\over {e^{p^0/T} -1}}\eqno(10)
$$

For our power counting rules, it is easiest to consider the
propagators with the time-orderings separated as in Eqs.(4,9a).
Suppose now that
these propagators are in a loop diagram, with integration over
momenta. The distributions
tell us that the average momentum is of the order of $T$. For each
diagram, we thus get $^9$:

1) $T^3$ corresponding to $d^3p$

2) $1/T$ for each propagator
from  the $1/2p^0$ factor

3) a power of $T$ for each factor of $p_\mu$ in the numerators
in vertices

4) for each propagator, other than the first, a power of
${1/k}$, where $k$ is the external momentum ($\sim gT$)

5) a factor $k/T$, one for each loop, for two or more
propagators of the same statistical type.

The last two rules can be understood as follows.
In using Eqs.(4,9a) for the propagators in any Feynman diagram,
we encounter many terms
corresponding to different ordering of the time arguments
$x^0, y^0, z^0, etc.$
An obvious strategy for simplification is to carry out the
time-integrations first, introducing convergence
factors $e^{\pm \epsilon x^0},e^{\pm \epsilon y^0} $ etc.,
$\epsilon$ small
and positive, as required. The integrations give
energy-denominators and
bring the result to a form where simplification due to
the external momenta
being soft $(\sim g T)$ relative to the loop momenta ($\sim T$)
can be implemented easily. Among energy-denominators, there will be
some of the form $ p^0-q^0 \pm k^0$, where $p,q$ refer to loop momenta.
Because their difference is involved, such a denominator is
of the order of $k$,
eventhough $p^0, ~q^0$ can be of the order of $T$. This gives the $1/k$-factor
of rule 4.
If we have only one propagator, time-integration gives only
an energy-conservation $\delta$-function; thus the $1/k$-factors are
only for
propagators other than the first.
The last rule has to do with the fact that the difference of statistical
distributions such as $n_p-n_q$ is what arises. With $p-q\approx k$,
this gives rule 5. The last two rules will become
clearer in the course of the evaluation of the two-point function
which we shall do shortly.

We can apply these rules, as an example, to the three-point
function and its first correction as shown.
Because of the derivative coupling, fig.(2a) goes as $gk\sim g^2T$.
For fig.(2b), first of all we have $g^3$, then
$T^3$ from rule 1, $1/T^3$ from rule
2, $T^3$ from rule 3, $1/k^2$ from rule 4 and $k/T$ from rule 5.
With $k\sim gT$, this is again $\sim g^2T$. Thus fig.(2b) must be
included in our calculation of $\Gamma [A]$. Continuing in a
similar way, one can see that all diagrams which should be included
in $\G$ are one-loop diagrams $^9$; further, only the $T$-dependent
terms of
these one-loop diagrams (and of course with external momenta small
compared to $T$) are important. Thermal one-loop diagrams with
loop momenta $\sim T$ and external momenta
${\buildrel < \over \sim}gT$ are called hard thermal loops.
\vskip .6cm
\noindent{\twelvebf 3. Calculation of the two-point function}
\vskip .2in
Let us now carry out an explicit calculation, say for the two-point
function. This may be a trifle tedious, but is well worth the effort
since it illustrates many of the features which generalize to
$n$-point functions $^1$.
We consider a one-loop quark graph with two external gluon
lines, in other words, the gluon polarization diagram.
The relevant part of the Lagrangian for the quark fields
$q, \bar q$ is

$$ {\cal L} = \bar q ~i \gamma \cdot ( \partial + A) ~q  \eqno(11)$$

\noindent where $A_{\mu} = -i t^a A^a_{\mu}$ is the
Lie-algebra-valued
gluon vector
potential, $t^a$ are hermitian matrices corresponding to the generators
of the Lie algebra in the representation to which the quarks
belong; $[t^a,t^b]=if^{abc}t^c$, where $f^{abc}$ are the structure
constants, and ${\rm Tr}(t^at^b)={1\over 2}\delta^{ab}$.
We take the gauge group to be $SU(N)$ with $N_F$ flavors of
quarks. From now on, we
shall not explicitly display the quark-gluon coupling constant $g$,
as it is easily recovered at any stage by $A \rightarrow g A$.
The one-loop quark graphs are given by the effective action
$$
\Gamma = -i ~\Tr ~\log ~(1 + S ~\gamma \cdot A)  \eqno(12)
$$
The two-gluon term in $\Gamma$ is given by
$$
\Gamma ^{(2)} = {i \over 2} \int d^4x d^4y~\Tr \left[ ~ \gA(x)
S(x,y)  \gA(y) S(y,x) \right] \eqno(13)
$$
In using Eq.(4) for the propagator, we find four terms
in $\Gamma^{(2)}$
with $x^0 > y^0 $ and four terms with $y^0 > x^0.$ Writing
$$
A_\mu (x) = \int {d^4k \over (2 \pi)^4} ~e^{ikx} ~A_\mu (k)
\eqno(14)
$$
and carrying out the time-integrations we get
$$
 \eqalignno {\Gamma^{(2)} = -{\textstyle {1 \over 2}}
 \int d\mu(k)\int {d^3q \over (2\pi)^3} &{1\over 2 p^0}{1\over 2 q^0} \left[
{}~T(p,q) \bigl( {\ap \bq \over p^0-q^0-k^0-i\epsilon}-
{\aq \bp \over p^0-q^0-k^0+i\epsilon} \bigr)+\right. \cr
&\left. T(p,q') \bigl( {\ap \abq \over p^0+q^0-k^0-i\epsilon}-
{\bp \bbq \over p^0+q^0-k^0+i\epsilon} \bigr)+ \right. \cr
&\left. T(p',q) \bigl( {\abp \aq \over p^0+q^0+k^0-i\epsilon}-
{\bbp \bq \over p^0+q^0+k^0+i\epsilon} \bigr)+ \right. \cr
&\left. T(p',q') \bigl( {\abp \bbq \over p^0-q^0+k^0-i\epsilon}-
{\bbp \abq \over p^0-q^0+k^0+i\epsilon} \bigr) \right]  &(15) \cr} $$

\noindent where
$$\eqalignno{
T(p,q) &= \Tr \lbrack ~\gA(k) \gp \gA(k') \gq ~\rbrack &(16)\cr
d\mu (k) &= (2\pi)^4 \delta^{(4)} (k+k'){d^4k \over (2\pi)^4}
             {d^4k' \over (2\pi)^4} &(17)\cr}
$$

In Eq.(15), $\vec p = \vec q + \vec k.$ Since $p^0 = |\vec q + \vec k |
\simeq q^0  + \vec q \cdot \vec k/q^0$ for $|\vec k|$ small
compared to $|\vec q|,$ the denominators in Eq.(15) involve $ k \cdot Q,
\; k \cdot Q'$ and $2 q^0 + k \cdot Q, \; 2 q^0 + k \cdot Q'$ where

$$ Q= (1, ~{\vec q \over q^0}), \; Q' = (1, ~-{\vec q \over q^0})
\eqno (18)$$
Notice that $Q^\mu, Q'^\mu $ are null vectors, $Q^\mu Q_\mu =Q'^\mu
Q'_\mu =0$. This is a remnant of the $\delta$-function in Eq.(8)
enforcing the mass-shell condition for the quarks.

The $i \epsilon$'s in the denominators in Eq.(15)
will be let go to zero at this stage.
The $i\epsilon$'s can contribute
to the imaginary part of the two-point function,
corresponding to the
Landau damping of the gluon field.
The retarded rather than time-ordered propagators are appropriate
for a discussion of damping effects; we shall obtain the correct
imaginary part by an appropriate continuation later. For the moment,
let us ignore the $\iep$'s.
Using $\alpha,\beta$
from Eq.(5), we find, for the temperature-dependent part of $\Gamma^{(2)},$

$$\eqalignno{\Gamma^{(2)} = -{\textstyle{1 \over 2}} \int
d\mu(k)\int {d^3q \over (2\pi)^3}&{1\over 2 p^0}{1 \over 2 q^0} ~\left[
(n_q-n_p){T(p,q) \over p^0-q^0-k^0}
+ (\bar n_q-\bar n_p){T(p',q') \over p^0-q^0+k^0}-\right. \cr
&\left. (n_p+\bar n_q){T(p,q') \over p^0+q^0-k^0}
-(\bar n_p+n_q){T(p',q) \over p^0+q^0+k^0} \right]. &(19) \cr}$$

We see that the result is linear in the distribution functions,
a property
which holds in general for the $n$-point functions.
Notice also how soft
denominators like $p^0-q^0 \pm k^0$ arise from the time-integration.
For $|\vec k|$ small compared to the loop momentum $|\vec q|,$
we can write
$$ p^0 -q^0 -k^0 \simeq - \kq ~~~~~~~~ p^0 -q^0 +k^0 \simeq  \kq'$$
$$ p^0 +q^0 \pm k^0 \simeq 2 q^0  \eqno (20)$$

$$T(p,q) \simeq 8 q^{0^2} \tr (\aoq \atq)$$
$$T(p',q') \simeq 8 q^{0^2} \tr (\aoq' \atq')   \eqno(21)$$
$$T(p',q) \simeq T(p,q') \simeq 4 q^{0^2} \tr (\aoq' \atq + \aoq \atq' -
2 A_1 \cdot A_2)$$

\noindent where $A_1 = A(k), \; A_2 = A(k') $ and the remaining
trace
in the expressions
for $T$'s, denoted by $\tr,$ is over color indices. The difference
of
distribution functions can also be approximated as $n_p - n_q \simeq {dn \over
dq^0} \vqk.$ (This is the extra factor of $k/T$ mentioned as
rule 5 above.)
Using these results, Eq.(19) simplifies to

$$\eqalignno{ \Gamma^{(2)} = -{\textstyle {1 \over 2}}
 \int d\mu(k)\int {d^3q \over (2\pi)^3}
{}~&\tr \Biggl[~ \bigl( {dn \over dq^0} {\aoq \atq \over \kq}-
{d\bar n \over dq^0} {\aoq' \atq' \over \kq'}\bigr) 2\vqk \cr
&\left. -{n +\bar n \over q^0}(\aoq' \atq + \aoq \atq' -
2 A_1 \cdot A_2) \right]  & (22) \cr } $$

We have the result

$$\int d^3q {dn \over dq^0} f(Q) = -\int d^3q {2n \over q^0} f(Q)
  \eqno(23)$$
for any function $f$ of $Q,$ or $Q'.$ We can further use $ 2\vqk = \kq' -\kq.$
Eq.(22) then simplifies to

$$\eqalignno{\Gamma^{(2)} = -{\textstyle {1 \over 2}} \int d\mu(k)\int
{d^3q \over (2\pi)^3}~\tr \Biggl[~
& {n +\bar n \over q^0}(2\aoq \atq - \aoq \atq' - \aoq' \atq +2 A_1\cdot
 A_2) \cr
&\left. -{n \over q^0} 2\aoq \atq {\kq' \over \kq} -
{\bar n \over q^0} 2\aoq' \atq' {\kq \over \kq'} \right]. &(24)\cr}
$$
The angular integration in Eq.(24) over the directions of $\vec q$
(or $\vec Q$)
can help in simplifying it further by virtue of

$$ \int d \Omega ~(2 \aoq \atq - \aoq \atq' - \aoq' \atq +2 A_1 A_2)
   =  \int d\Omega~ (2 \aoq \atq')  \eqno(25)$$
Defining

$$A_+ = {A\dq \over 2},~~~~~~~~ A_- = {A\dq' \over 2}  \eqno(26) $$

\noindent we can write Eq.(24) as

$$\Gamma^{(2)} = -{\textstyle {1 \over 2}} \int d\mu(k)
\int {d^3q \over (2\pi)^3}
{1 \over 2 q^0} ~16 ~\tr \left[~ A_{1+}A_{2-} (n+ \bar n) -
n {\kq' \over \kq} A_{1+}A_{2+}-
\bar n {\kq \over \kq'} A_{1-}A_{2-} \right].  \eqno(27) $$

This expression becomes more transparent when written in
coordinate space and finally with a Wick rotation to Euclidean
space. Let us define the Green's functions

$$G(x_1, x_2)=\int {e^{-ip \cdot (x_1-x_2)} \over p \cdot Q}
{d^4p \over (2\pi)^4}$$
$$G'(x_1, x_2)=\int {e^{-ip \cdot (x_1-x_2)} \over p \cdot Q'}
{d^4p \over (2\pi)^4}    \eqno(28)$$
In terms of the null vectors $Q = (1, \vec q/q^0), \;
Q' = (1,-\vec q/q^0), $ we can introduce the lightcone coordinates
$(u,v,x^T)$ as

$$u = {Q' \cdot x \over 2} , ~~~v = {Q \cdot x \over 2},~~~
\vec Q \cdot \vec x^T = 0  \eqno(29) $$

\noindent where $\vec Q = \vec q/q^0. $ We then have $Q \cdot \partial =
\partial_u,
\; Q' \cdot \partial = \partial_v.$ We shall also introduce a
Euclidean version of our results by the correspondence

$$ 2u \leftrightarrow z, ~~~~~2v \leftrightarrow \bar z $$
$$\partial_u =  Q \cdot \partial \leftrightarrow 2 \partial_z,~~~~
\partial_v =  Q' \cdot \partial \leftrightarrow 2 \partial_{\bar z}
 \eqno(30)$$
The Green's functions in Eq.(28) are the continuations of the Euclidean
functions

$$G_E(x_1,x_2) = {1 \over 2\pi i} {\delta^{(2)}(x_1^T-x_2^T)  \over
(\bar z_1-\bar z_2)}$$
$$G'_E(x_1,x_2) = {1 \over 2\pi i} {\delta^{(2)}(x_1^T-x_2^T)  \over
( z_1-z_2)}    \eqno (31)$$
More precisely, the Green's functions in Eq.(28) obey the equations
$Q\cdot \partial G(x_1, x_2)= -i\delta (x_1-x_2),~ Q'\cdot \partial
G'(x_1, x_2)=-i\delta (x_1-x_2)$. The Green's functions in Eq.(31)
are the
solutions to the corresponding Euclidean equations $2\partial_z G_E
(x_1-x_2)=-i\delta (x_1-x_2),~2\partial_{\bar z}G'_E (x_1-x_2)=-i\delta
(x_1-x_2)$.
This leads to the correspondence

$$ {\kq' \over \kq} \longleftrightarrow {1 \over \pi}~{1 \over \bar z_{12} \bar
z_{21}}$$
$$ {\kq \over \kq'} \longleftrightarrow {1 \over \pi}~{1 \over z_{12} z_{21}}$$
$$ {1\over \kq} \longleftrightarrow {1 \over  2 \pi i}~{1 \over \bar z_{12}}
 \eqno(32)$$

\noindent where $z_{ij} = z_i -z_j,$ etc. Eq.(27) for $\Gamma^{(2)}$ can
then be written as

$$\eqalignno{\Gamma^{(2)} &= -{\textstyle {1 \over 2}}
\int {d^3q \over (2\pi)^3}
{1 \over 2q^0} ~16
{}~\tr \left[ ~(n + \bar n)~\int d^4x~A_+(x)A_-(x) \right.\cr
&\left. -n ~\pi \int {d^2x^T}{d^2z_1\over \pi} {d^2z_2\over \pi}
{A_+(x_1)A_+(x_2) \over \bar z_{12}\bar z_{21}}-
\bar n ~\pi \int {d^2x^T}{d^2z_1\over \pi}{d^2z_2\over \pi}
{A_-(x_1)A_-(x_2) \over z_{12}z_{21}}
\right]. & (33)\cr} $$
This is finally starting to look nice. Let us
define the functional $I(A_+)$ by the formula
$$\eqalignno{ I(A_+) &= i \sum {(-1)^n \over n}\int ~
d^2x^T~{d^2z_1\over \pi}\cdots {d^2z_n \over \pi}~{\tr(A_+(x_1)
 \cdots A_+(x_n))
\over \bar z_{12}  \bar z_{23} \cdots \bar z_{n1} }
\cr
&= {i\over 2} \int ~{d^2x^T}~{d^2z_1 \over \pi}
{d^2z_2 \over \pi} ~{\tr(A_+(x_1)A_+(x_2)) \over
\bar z_{12} \bar z_{21} }
 + \cdots &(34) \cr}$$
We shall show later that $I(A_+)$ is related to the eikonal for a
Chern-Simons theory. Eq.(33) for $\Gamma^{(2)}$ can finally
be written as
$$\Gamma^{(2)} = \int {d^3q \over (2\pi)^3}{1 \over 2q^0}
                   ~~K^{(2)}[A_+,A_-]     \eqno(35a)$$

\noindent where
$$ K[A_+,A_-] =-16 ~\left[~ {(n +\bar n)\over 2} \int d^4x~\tr ~\bigl(
A_+(x)A_-(x)\bigr)
+n ~i \pi  I(A_+) + \bar n ~i \pi \tilde I(A_-) \right]   \eqno(35b)$$

\noindent $K^{(2)}[A_+,A_-]$ in Eq.(35a) denotes terms in $K$
which are quadratic
in $A.$
$\tilde I $ is obtained from $I$ by $z \leftrightarrow \bar z.$

Although we have introduced different distribution functions $n, \bar n$
for quarks and antiquarks, it is only $n + \bar n$ which is relevant at
high temperatures. We see that, by virtue of $\int d\Omega ~I(A_+) =
\int d\Omega ~\tilde I(A_-),$ we can write Eq.(35b) as

$$ K[A_+,A_-] =-16 ~{n + \bar n \over 2}~\left[~ \int d^4x
{}~\tr \bigl( A_+(x)A_-(x) \bigr)
+i \pi I(A_+) + i \pi \tilde I(A_-)  \right] . \eqno(36)$$

The integral over the magnitude of $\vec q$ in Eqs.(35) can be easily
carried out. With zero chemical potential,
$$\Gamma^{(2)} = {-T^2 \over 12\pi} \int d\Omega ~\left[~ \int
d^4x~\tr \bigl( A_+ ~A_- \bigr)
+i \pi I^{(2)}(A_+) + i \pi \tilde I^{(2)}(A_-)  \right]. \eqno(37)$$

\vskip .6cm
\noindent{\twelvebf 4. Properties of $\Gamma [A]$}
\vskip .2in
We can extract certain properties of $\Gamma [A]$ from the
power counting rules discussed earlier. There are two key properties
which are important to our analysis.

1) $\Gamma [A]$ is gauge-invariant with respect to gauge transformations
of the gauge potential $A_\mu$ and is independent of the gauge-fixing
used to define the gluon propagators $^{9,10}$.

We can understand how the kinematics of hard thermal loops can lead to
gauge invariance. The thermal propagators satisfy the same
differential equations as the zero temperature propagators.
This is evident from Eqs.(8,9b); only the choice of the homogeneous
solution, equivalently boundary condition, is different. As a result,
the generating functional of one-particle irreducible
vertices, viz., $\Gamma [A,c,{\bar c}, q, {\bar q}]$ obeys the standard
BRST Ward identities. $c$ and $\bar c$ are the ghost and antighost
fields respectively.
For the usual gauge fixing term $\lambda (\partial
\cdot A)^2$, we thus have $^{11}$
$$
\int d^4x~ \left[ {{\delta \G}\over {\delta A_\mu}} {{\delta \G}\over
{\delta K_\mu}}~+~{{\delta \G}\over {\delta c}}{{\delta \G}\over {\delta L}}
{}~-\lambda (\partial \cdot A) {{\delta \G}\over {\delta {\bar c}}}
\right] ~=0 \eqno(38)
$$
$K_\mu$ and $L$ are sources, corresponding to the terms
$K^a_\mu (D_\mu c)^a,~ f^{abc}L^a c^bc^c$ added to the action.
In the hard thermal loop approximation, terms involving the ghosts
are subdominant. Recall that the ghost-gluon coupling involves
$f^{abc} A^{a\mu} (\partial_\mu {\bar c}^b)c^c$; the derivative is on the
antighost field. In a diagram with external ghosts, this is a power
of external momentum and therefore such a diagram is smaller compared
to a similar diagram with the ghost lines replaced by gluons where
there are contributions with all derivatives on the internal lines
which give powers of $T$. (See
figs.(3).) The terms ${{\delta \G}\over {\delta c}},~{{\delta \G}\over
{\delta {\bar c}}}$ in Eq.(38) are thus negligible in the hard thermal loop
approximation. Further, we are interested only in one-loop terms.
The identity (38) then gives the gauge invariance of $\G$.
Thus effectively, in a high-$T$-expansion, the leading term $\G [A]$
which is proportional to $T^2$ is gauge-invariant. Notice also that
since the thermal contribution to the propagator is on-shell, the
$T$-dependent part of a
one-loop diagram is classical and so it is not surprising that the
BRST Ward identities reduce to the statement of gauge invariance.
The fact that
$\G$ does not depend on the gauge choice for the gluon propagators can be
seen by similar arguments; identities for the variation of $\G$
for changes in $\lambda$ can be written down and simplified
$^{9,10}$.

2) $\Gamma [A]$ has the form $^{12}$
$$
\Gamma=
(N+ {\textstyle {1\over 2}}N_F)~{T^2\over 12\pi}~\left[\int d^4x~
2\pi A_0^aA_0^a
+\int d\Omega ~W(A\cdot Q)\right]. \eqno(39)
$$
Here $Q_\mu$ is the null vector $(1, \vQ )$
with $\vQ^2=1,~Q_\mu Q_\mu=0$. The $A^a_0A^a_0$-term is the lowest
order Debye screening effect; it is clearly an electrostatic mass
term. The key point about Eq.(39) is that, for each $\vQ$,
$\G [A]$ involves
essentially only two components of the gauge potential, $A_0$ and
$A\cdot Q$. The $d\Omega$-integration will bring in all components
of the potential, but for each $\vQ$ only
two components are needed.
We have already seen this structure
in the explicit calculation of the two-point function, where
$\vQ$ was the angular part of the loop-momentum $\vq$. The
$d\Omega$-integration in Eq.(39) is over the orientations of $\vQ$
and is the unfinished part of the loop integration, after the integration
over the modulus of the loop momentum has been done.
The structure of Eq.(39) can be seen by analysis of diagrams again.
For example, for the diagram of fig.(3b), with the derivative gluon
coupling, since $p\cdot A \approx
q\cdot A$, the possible tensor structures are $q^2 A^2$ and
$(q\cdot A)^2$. The former is zero since the propagator involves the
$\delta$-function $\delta (q^2)$. Writing $q= \vert \vq \vert (1, \vQ )$
and carrying out the $\vert \vq \vert$-integration, we are left with
a structure like Eq.(39). This argument generalizes to diagrams with
arbitrary number of external gluons.

Given these two properties of $\G$ one can determine $W$ and
hence $\G$ simply by the requirement of gauge invariance $^{12}$.
The condition for gauge invariance of $\Gamma$ is
$$
\int d\Omega~\delta W=4\pi\int d^4x ~{\dot A}_0^a \omega^a\eqno(40)
$$
\noindent where ${\dot A}^a_0$ is the time-derivative of $A_0^a$
and $\omega=-it^a\omega^a$ is the parameter of the gauge
transformation, i.e., $\delta A_\mu=\partial_\mu \omega
+[A_\mu,\omega]$. Eq.(40) can be realized by
$$
\delta W=\int d^4x ~{\dot A}^a\cdot Q \omega^a.\eqno(41)
$$
\noindent One can check that Eq.(41) is indeed the way gauge invariance
is realized by analysis of the diagrams. It is clearly so for
the two-point function from our explicit calculation.
We now rewrite Eq.(41), using
$$\delta W=-\int d^4x(Q\cdot\partial{\delta W\over \delta(A\cdot Q)}
+[A\cdot Q,{\delta W\over\delta(A\cdot Q)}])^a\omega^a\eqno(42)
$$
\noindent as
$$
{\partial f\over\partial u}+[A\cdot Q,f]+{1\over 2}
{\partial(A\cdot Q)\over \partial v}=0,\eqno(43)
$$
\noindent where
$$
f={\delta W\over\delta(A\cdot Q)}+
{\textstyle{1\over 2}}A\cdot Q.\eqno(44)
$$

\noindent We have used the lightcone coordinates from Eq.(29).
We now make the Wick rotation as in Eq.(30) which gives
$A_{+}= {\textstyle{1\over 2}}A\cdot Q~\rightarrow A_z$;
also rename $f$ as
$$
a_{\bar z}=-f
=-{1\over2}{\delta W\over\delta A_z}-A_z\eqno(45)
$$
The condition of gauge invariance, Eq.(43),
then becomes
$$
\partial_{\bar z}A_z-\partial_z a_{\bar z}+[a_{\bar z},A_z]=0.
\eqno(46)
$$

If $A_z$, $a_{\bar z}$ are thought of as the gauge
potentials of another gauge theory, we see that
Eq.(46) is the vanishing of the field strength or
curvature $F_{z{\bar z}}$. The gauge theory whose
equations of motion say that the field strengths vanish is
the Chern-Simons theory.
We shall therefore put aside Eq.(46) for a moment and turn to
a short digression on the
Chern-Simons theory, returning to Eq.(46) and its solution in
section 6. Of course, an understanding of
Chern-Simons theory is not
absolutely essential to solving Eq.(46).  One can simply
solve Eq.(46) and regard Chern-Simons theory as an interpretation of
the mathematical steps along the way. However Chern-Simons
theory does illuminate many of the nice geometrical properties
of the final result and is a worthwhile
digression.
\vskip .6cm
\noindent{\twelvebf 5. Chern-Simons and WZNW Theories}
\vskip .2in
The Chern-Simons theory is a gauge theory in two
space (and one time) dimensions $^{13,14}$. The action is given by
$$
S={\k\over 4\pi}\int_{M\times[t_i,t_f]}d^3x~\epsilon^{\mu\nu\alpha}
{}~\tr(a_\mu\partial_\nu
a_\alpha+{\textstyle{2\over3}}a_\mu a_\nu a_\alpha).
\eqno(47)
$$
\noindent Here $a_\mu$ is the Lie algebra valued gauge potential,
$a_\mu=-it^a a_\mu^a$.
$\k$ is a constant whose precise value we do not need to specify
at this stage. We shall consider the spatial manifold to
be ${\bf R}^2$, or ${\bf C}$ since we shall be using complex
coordinates $z=x+iy$, ${\bar z}=x-iy$.
(Actually, we have sufficient regularity conditions at
spatial infinity that we may take $M$ to
be the Riemann sphere.) The equations of motion for the theory are
$$
F_{\mu\nu}= \partial_\mu a_\nu - \partial_\nu a_\mu +[a_\mu ,a_\nu ]
=0.\eqno(48)
$$

The theory is best analyzed, for our purposes, in the
gauge where $a_0$ is set to zero. In this gauge, the
equations of motion (48) tell us that $a_z$, $a_{\bar z}$
are independent of time, but must satisfy the constraint
$$
F_{{\bar z}z}\equiv\partial_{\bar z}a_z-
\partial_z a_{\bar z}+[a_{\bar z},a_z]=0.\eqno(49)
$$
\noindent This constraint is just the Gauss law of the CS
gauge theory. It can be solved for $a_{\bar z}$ as a function
of $a_z$, at least as a power series in $a_z$. The result
is
$$
a_{\bar z}=\sum(-1)^{n-1}
\int {d^2z_1\over\pi}\cdots{d^2z_n\over\pi}~
{a_z(z_1,{\bar z}_1)a_z(z_2,{\bar z}_2)\ldots
a_z(z_n,{\bar z}_n)\over
({\bar z}-{\bar z}_1)({\bar z}_1-{\bar z}_2)\ldots({\bar z}_n-{\bar z})}.
\eqno(50)
$$
\noindent This can be easily checked using $\partial_z ({1\over {\bar z}-
{\bar z}'})=\pi\delta^{(2)}(z-z')$.

In the $a_0=0$ gauge, the action becomes
$$
S={i\k\over\pi}\int dt d^2x ~\tr (a_{\bar z}\partial_0 a_z).
\eqno(51)
$$
\noindent This shows that $a_{\bar z}$ is essentially canonically
conjugate to $a_z$. In fact in carrying out a variation
of $S$, we find the surface term $\theta(t_f)-\theta(t_i)$, where
$$
\theta={i\k\over\pi}\int_Md^2x~\tr(a_{\bar z}\delta a_z).\eqno(52)
$$
\noindent (We assume $a_{\bar z}\delta a_z$ to vanish at
spatial infinity.)
$\theta$ is the canonical one-form of the CS theory. (This is so by
definition; the canonical one-form in any theory can be defined by
$\theta$, where the surface term in the variation of the action is
$\theta_f ~-~\theta_i$, the subscripts referring to the final and
initial data surfaces.)
$\theta$ is the analogue of $p_idx^i$ of point-particle mechanics; for an
action $S=\int dt dx~[{m{\dot x}^2\over2}-V(x)],$ we would find
$\theta=m{\dot x}_i\delta x^i=p_i\delta x^i$. We can make
another variation of $\theta$, antisymmetrized with respect to the
variation $\delta a_z$, denoted by the wedge product sign,
and write
$$\eqalignno{\omega \equiv \delta\theta&={i\k\over\pi}
\int_Md^2x ~\tr(\delta a_{\bar z}\wedge\delta a_z)\cr
&={\textstyle{1\over2}}\int d^2x d^2x'
{}~\omega_{AB}(x,x')~\delta\xi^A(x)\wedge\delta\xi^B(x')
&(53)\cr}
$$
\noindent where
$$
\omega_{AB}(x,x')=-{i\k\over 2\pi}\left(\matrix{
0&\delta(x-x')\delta^{ab}\cr
-\delta(x-x')\delta^{ab}&0\cr}\right)\eqno(54)
$$
\noindent and $\delta\xi^A=(\delta a_{\bar z}^a,\delta a_z^a)$.
$\omega$ defined by Eqs.(53,54) is called the
symplectic structure and is of course the analogue of
$dp_idx^i$ of particle mechanics. The inverse of $\omega_{AB}$ gives
the Poisson brackets, the commutators being $i$ times the Poisson
brackets.
For our case we get
$$[\xi^A(x),\xi^B(x')]=i(\omega^{-1})^{AB}(x,x')$$
\noindent or
$$[a^a_{\bar z}(x),a_z^b(x)]={2\pi\over \k}\delta^{ab}\delta^{(2)}(x-x').
\eqno(55)
$$
\noindent One does not have to go through $\theta$ and $\omega$
to arrive at Eq.(55). One could simply use the fact that, from
Eq.(51) the canonical momenta are $\pi=a_{\bar z}$, ${\bar \pi}=0$.
This is thus a constrained system in the Dirac sense and using the
theory of constraints one can derive Eq.(55).
The procedure of using
$\omega_{AB}(x,x')$ is quicker.

In the expression (52) for $\theta$, $a_{\bar z}$ is independent
of $a_z$. We can however express $a_{\bar z}$ as a function of
$a_z$ via the constraint Eq.(49) or equivalently Eq.(50) and functionally
integrate $\theta$. In other words, we define $I(a_z)$ by
$$
\delta I={i\k \over\pi}\int d^2x~\tr
[a_{\bar z}(a_z)\delta a_z].\eqno(56)
$$
\noindent The solution for $I$ is given by
$$
I=i\k\sum{(-1)^n\over n}\int {d^2z_1\over\pi}\cdots
{d^2z_n\over\pi}~
{\tr(a_z(z_1,{\bar z}_1)\ldots a_z(z_n,{\bar z}_n))\over
{\bar z}_{12}{\bar z}_{23}\ldots{\bar z}_{n-1n}{\bar z}_{n1}}.
\eqno(57)
$$

The quantity $I$ has a rather simple interpretation. For
one-dimensional point-particle mechanics, $\theta$, as
we mentioned earlier, is given by $pdx$. $p$ is independent
of $x$ to begin with, but we can express it as a function of
$x$ via a constraint such as of fixed energy, e.g.,
${p^2\over2m}+V(x)=E$. Integral of
$\theta=pdx$ then gives Hamilton's principal function or the
eikonal, familiar as the exponent for the WKB wave functions
of one-dimensional quantum mechanics. We have an
analogous situation with Eq.(49) expressing $a_{\bar z}$ as a
function of $a_z$. $I$ is thus an eikonal of the CS
theory $^{14,15}$.

$I$ is in fact the Wess-Zumino-Novikov-Witten (WZNW)
action $^{16}$. We can write the
gauge potential $a_z$ as $a_z=-\partial_zUU^{-1}$ where $U$ is
in general not unitary; it is an $SL(N,{\bf C})$ matrix for
gauge group $SU(N)$. Notice that since $\partial_z$ has
an inverse by virtue of $\partial_z{1\over ({\bar z}-{\bar z}')} =\pi
\delta^{(2)}(z - z')$,
such a $U$ can be constructed for any $a_z$, at least as a power
series in $a_z$. $I$ can then be written as $I=-i\k S_{WZNW}(U)$
where
$$
S_{WZNW}(U)={1\over2\pi}\int_Md^2x~\tr(\partial_zU\partial_{\bar z}
U^{-1})-{i\over 12\pi}
\int_{M^3}d^3x~\epsilon^{\mu\nu\alpha}\tr(U^{-1}\partial_\mu U
U^{-1}\partial_\nu U U^{-1}\partial_\alpha U).\eqno(58)
$$
\noindent The second term, the so-called Wess-Zumino (WZ) term,
involves an extension of $U$ into
a three-dimensional space. We take $M^3=M\times [0,1]$ with
$U(z,{\bar z},0)=1$, $U(z,{\bar z},1)=U(z,{\bar z})$. The integrand of the
WZ term becomes a total derivative once a (local)
parametrization is chosen for $U$. This ensures that the physics
is independent of
small changes in how one makes
the extension of $U$ into the extra dimension.
If we have two globally
inequivalent extensions, say $U_1$ and $U_2$, the difference
$S_{WZNW}(U_1)-S_{WZNW}(U_2) =2\pi Q[g]$ where $Q[g]$ is the winding number
of the map $g: S^3 \rightarrow G $; $G$ is the space in which the
matrices $U$ take values, $SL(N,{\bf C})$ in the present case and
$g = U_1$ on the upper hemisphere of $S^3$, which can be taken as
one copy of $M^3$ and $g=U_2$ on the lower hemisphere,
taken as another copy of $M^3$. Since $Q$ is an integer,
$e^{i\k S_{WZNW}}$ is independent of globally different extensions
if $\k$ is an
integer $^{16}$.

The WZNW-action also obeys the following property,
sometimes called the Polyakov-Wiegmann formula $^{17}$.
$$
S(hU)~=~ S(h)~+~S(U)~-~ {1\over \pi} \int_{M^2}~ {\rm
Tr}(h^{-1}\partial_{\bar z}h~\partial_{z}U~U^{-1})\eqno(59)
$$
The crucial point here is that in the cross-term $h$ has only
${\bar z}$-derivative and $U$ has only $z$-derivative. (They become
$\partial_-,~\partial_+$ in Minkowski space.)

The relationship of $S_{WZNW}$
to $I$ is easily seen by considering variations. Under
the variation $U\rightarrow e^\varphi U\simeq(1+\varphi)U$,
we find $\delta a_z=-D_z\varphi=-(\partial_z\varphi+
[a_z,\varphi])$ and
$$
\delta S_{WZNW}={1\over\pi}\int d^2x~\tr(\partial_{\bar z}\varphi a_z).
\eqno(60)
$$
\noindent Partially integrating and using $F_{{\bar z}z}=0$
and $\delta a_z=-D_z\varphi$, we find that
$S_{WZNW}$ obeys Eq.(56) except for a factor $(-i\k )$, thus
identifying $I=-i\k S_{WZNW}$.

Another quantity of interest is
$$
K=-{1\over \pi}\left[~\k \int d^2x~ \tr (a_{\bar z} a_z)
+i\pi I(a_z) +i\pi {\tilde I}(a_{\bar z})\right].\eqno(61)
$$
It is easily checked that this is gauge-invariant. $K$ has a nice
interpretation as a K\"ahler potential,
which I shall not go into here $^1$.

Finally notice that the eikonal $I$ of Eq.(57) may be regarded
as the expansion in powers of $a_z$ of the logarithm of
the functional determinant of $D_z=\partial_z+a_z$; i.e.,
 $I=(-i\k )\log\det
D_z=-i\k ~\Tr\log D_z$. The expression $K$ of Eq.(61)
is then given by $-\k \Tr \log(D_zD_{\bar z})$.
The expansion of this expression in powers of the potential
obviously gives $I$ and ${\tilde I}$. The extra term
$\int{1\over \pi}\tr(a_{\bar z}a_z)$ is precisely the local counterterm
needed to give a gauge invariantly regulated meaning to
$\Tr\log(D_zD_{\bar z})$.
\vskip .6cm
\noindent{\twelvebf 6. Solution for $\G [A]$}
\vskip .2in
We now return to Eqs.(45,46) for the quark-gluon plasma.
We can rewrite Eq.(45) as
$$
\delta W=4\int d^4x ~\tr(a_{\bar z}\delta A_z)- \delta
\int d^4x~ A_z^a A_z^a .
\eqno(62)
$$
\noindent Comparing with Eq.(56), we see that, since
$a_{\bar z}$, $A_z$ obey
the constraint Eq.(46), the solution is related to the eikonal $I$.
The difference here is that $A_z$ and $a_{\bar z}$ depend on all four
coordinates $x_\mu$, not just $z$, ${\bar z}$. However, there are no
derivatives with respect to the transverse coordinates $x^T$ in
Eq.(46)
and hence the solution for $a_{\bar z}$ in terms of $a_z$ is the
same as in Eq.(50), with $a_z$ depending on $x^T$ in addition to
$z$, ${\bar z}$. The $x^T$-argument of all $a_z$ factors is the
same, i.e.,
$$
a_{\bar z} = \sum (-1)^{n-1}\int {d^2z_1\over\pi}
\cdots{d^2z_n\over\pi}
{}~{A_z(z_1,{\bar z}_1,x^T)A_z(z_2,{\bar z}_2,x^T)
\ldots A_z(z_n,{\bar z}_n,x^T)
\over ({\bar z}-{\bar z}_1){\bar z}_{12}\ldots {\bar z}_{n-1n}
({\bar z}_n-{\bar z})}.
\eqno(63)
$$
\noindent For the eikonal we get the same expression as Eq.(57)
but with
integration over the transverse coordinates, i.e.,
$$
I=i\k \sum {(-1)^n\over n} \int d^2x^T~
{d^2z_1\over\pi}\ldots {d^2z_n\over\pi}~
{\tr (A_z(x_1)\ldots A_z(x_n))\over {\bar z}_{12}{\bar z}_{23}
\ldots
{\bar z}_{n1}}.\eqno(64)
$$
\noindent Since it is not relevant to the present discussion, we have,
for the moment, set $\k =1$. We then find the solution to Eq.(62) as
$$
W=-4\pi i I(A_z)-\int d^4x~ A_z^a A_z^a .\eqno(65)
$$
(Strictly speaking $a_{\bar z}$ and $A_z$ are not complex
conjugates; the
analogy holds better with a Chern-Simons theory of complex gauge
group.
However, we are only using the Chern-Simons analogy to obtain
Eq.(65).
The expression for $\Gamma [A]$ can also be directly checked to be
a solution to Eqs.(45,46).)

\noindent From Eq.(39), $\Gamma$ is given by
$$
\Gamma=(N+ {\textstyle {1\over 2}}N_F){T^2\over 12\pi}
\left[\int d^4x~ 2\pi
A_0^aA_0^a -\int d\Omega \{\int d^4x~~A_{z}^a A_{z}^a +
4\pi i I(A_{z})\}\right].
\eqno(66)
$$
\noindent We can now use the identity
$$
\int d\Omega ~\tr(A_zA_{\bar z}) =
-{\textstyle{1 \over 2}}~\left[~2\pi A_0^a A_0^a -
\int d\Omega ~(A_z^aA_z^a) \right], \eqno(67)
$$
which follows by straightforward $d\Omega$-integration, to
write
$$\eqalignno{\Gamma&=-(N+{\textstyle {1\over 2}}N_F)
{T^2\over 6\pi}
\int d\Omega
\left[\int d^4x~\tr(A_{z}A_{\bar z}) +i\pi I(A_{z}) +
i\pi \tilde I(A_{\bar z})
\right]&(68a)\cr
&=(N+{\textstyle {1\over 2}}N_F){T^2\over 6}\int d\Omega
{}~K(A_{z}, A_{\bar z})&(68b)\cr}
$$
\noindent where $K$ is given by Eq.(61) with $\k =1$ and with
the additional integration over
coordinates ${\vec x}^T$ transverse to ${\vec Q}$.
If we write $A_{z}=
-\partial_z U U^{-1}$ and
$A_{\bar z}={U^{\dagger}}^{-1}\partial_{\bar z} U$,
we can write Eq.(68) in terms of $S_{WZNW}$, using Eq.(59), as
$$
\Gamma = -(N+{\textstyle {1\over 2}}N_F){T^2\over 6}S_{WZNW}(U^{\dagger}U).
\eqno(69)
$$
\noindent (Again, a suitable additional integration over the transverse
coordinates is understood.)

 From Eqs.(41,42), it may seem that our solution
is ambiguous up to the addition of a purely gauge invariant
term. But for hard thermal loops, the additional structure that $W$
depends only on $A_z$ tells us that the gauge
invariant piece must obey $D_z{\delta W\over \delta A_z}=0$.
Since $\partial_z$ is invertible, at least
perturbatively, there is no nontrivial solution to
this equation. Thus Eq.(66) or Eq.(68) is the unique solution and
$\Gamma$ so
defined must indeed be the generator of hard thermal loops.

In the last section, we noted that $K(A_z , A_{\bar z} )$ can be considered as
$\Tr {\rm log}(D_z D_{\bar z})$. Since $D_z$ and $D_{\bar z}$ are the
chiral Dirac operators in two dimensions, $\Tr {\rm log}(D_z D_{\bar
z})$ is clearly the photon mass term of the Schwinger model, for Abelian
gauge fields. More generally, for non-Abelian fields as well, we can
consider $\Tr {\rm log}(D_z D_{\bar z})$ as a gauge-invariant mass term.
It is perhaps fitting that the gauge-invariant Debye screening mass term
in four-dimensional QCD is given by suitable integrations of such a
two-dimensional mass term (with, of course, the additional
$x^T$-dependence). Equally appropriately, the Chern-Simons term, from which we
derive $K(A_{z},A_{\bar z})$, is also the
mass term for gauge fields in three dimensions.

We close this section with two remarks on $\Gamma$. The
CS theory of Eq.(47) violates parity. We see that this parity violation
disappears, as indeed it should, by integration over
the orientations of ${\vec Q}$, for the QCD case.
Alternatively, expression (68a)
is manifestly parity-symmetric with $A_z \leftrightarrow A_{\bar z}$
(or $A_{+}\leftrightarrow A_{-}$ in Minkowski space),
$Q\leftrightarrow Q'$ under parity. Secondly, instead of setting
$\k =1$
and then having a prefactor $(N+{1\over 2}N_F){T^2\over 6}$ in
Eq.(68), we could simply choose $\k =(N+{1\over 2}N_F){T^2\over 6}$.
For WZNW actions, if $U$ has a non-Abelian
unitary part, any action we use must be an integer times
$S_{WZNW}$, as we have already seen.
In our case, $\Gamma$ involves $S_{WZNW}(U^{\dagger}U)$ or
$S_{WZNW}(H)$ where $H$ is hermitian. The third homotopy group
is trivial for the space of hermitian
matrices and so, as expected, there is no
argument for quantization of the coefficient, which
is $(N+{1\over 2}N_F){T^2\over 6}$ for us.
\vskip .6cm
\noindent{\twelvebf 7. Continuation to Minkowski space}
\vskip .2in
We shall now consider how the above results can be continued back to
Minkowski space $^2$. Evidently, we expect $A_z \rightarrow A_+,~
A_{\bar z}\rightarrow A_-$ and $\partial_z \rightarrow \partial_+,~
\partial_{\bar z} \rightarrow \partial_-$. The inverses of
$\partial_{\pm}$, however, require specification of
boundary conditions or $\iep$-prescriptions.
This can be easily seen in momentum space where the inverses
have denominators like $k\cdot Q$ which can vanish; one must give a
prescription on how these singularities are to be handled for the
$k^0$-integration. The appropriate $\iep$-prescription depends on
the physical context and the
quantity being considered. Very often, one is interested in the evolution
of fields in a plasma, which can be described by the operator field
equations. For this case, the retarded Green's functions are the
appropriate ones. Let me illustrate this with a simple example,
viz.,  a plasma in electrodynamics. The field equations can be written
as $^{18}$
$$
\partial^2 A^\mu ~=~
iS^{-1}{{\delta S}\over {\delta A_\mu ^{in}}}\equiv J^\mu \eqno(70)
$$
(We have chosen the Feynman gauge for simplicity.) $S$ is the scattering
operator
considered as a function of the incoming field $A_\mu^{in}$. We can
expand the current as a function of the field $A_\mu^{in}$, obtaining
to linear order in the field,
$$
J^\mu = j^\mu -i\int d^4y~ \Theta (x^0-y^0) [j^\mu (x),
j^\nu (y)] A_\nu^{in}(y) \eqno(71)
$$
where $j^\mu$ is the current in the interaction picture, say
${\bar q}_I \gamma^\mu q_I$ in terms of the fernion field $q_I$ in the
interaction picture. Notice that the we have the retarded commutator
of the currents $j^\mu$.
Eq.(71) is an operator equation which is generally true.
We can take any kind of expectation value of this equation;
the thermal result which is the
induced current in the plasma
is obtained by taking a thermal average.
(The result is the Kubo formula.)
To the order we are interested in, the incoming and interacting fields
can be considered the same and Eq.(70) gives its evolution.
The appropriate boundary condition for the
induced current which governs the evolution of the fields, we see,
is the retarded condition. Of course, the induced current
can also be
calculated in terms of the induced or effective action $\G$
as $-{{\delta \G}\over {\delta A_\mu}}$. The strategy for
continuation to Minkowski space is thus to calculate the induced current
in Euclidean space and then to use $\iep$'s appropriately in
denominators involving $k \cdot Q$'s so as to get the retarded
condition. If $J^\mu$ is expanded to higher orders in the
field, we encounter multiple retarded commutators. The $\iep$'s must be
inserted appropriately so that we get this retardation structure.
With this understanding, we can write the equations for the
evolution of fields of soft momenta in the quark-gluon plasma as
$$
D_\nu F^{\nu\mu,a}= J^{\mu,a} \eqno(72a)
$$
$$
J^{\mu,a}= \sum_1^\infty \int {d^4k_1\over (2\pi )^4}\cdots {d^4k_n\over
(2\pi)^4} e^{i(\sum k)\cdot x} J^{\mu,a}_n (k) \eqno(72b)
$$
$$\eqalignno{
J^{\mu,a}_n (k)= {\k \over \pi} &\int d\Omega \Biggl[ {\rm Tr}
\left( ({-it^a Q^\mu
\over 2}) A_- (k_1)+ A_+ (k_1) ({-it^a Q'^\mu \over 2})\right)\delta_{n,1}\cr
&+\left\{ -(2i)^{n-1}{\rm Tr} \left( ({-it^a Q^\mu\over 2})A_+(k_1)
\cdots A_+(k_n)
\right) F(k_1, \cdots k_n)+(Q\leftrightarrow Q')\right\} \Biggr]
&(72c)\cr}$$
where
$$\eqalignno{
F(k_1,\cdots k_n)&= \sum_{i=0}^n {{q_i}\over
{({\bar q}_i -{\bar q}_0)
({\bar q}_i-{\bar q}_1)\cdots ({\bar q}_i- {\bar q}_{i-1})
({\bar q}_i-{\bar q}_{i+1} )\cdots ({\bar q}_i-{\bar q}_n )}}&(73a)\cr
{\bar q}_i &= \sum_{j=1}^i (k_j \cdot Q -i\epsilon_j),
{}~~~~~~~~~~~q_i = \sum_{j=1}^i k_j\cdot Q' &(73b)\cr}
$$
We may reexpress the current as
$$
J^{\nu a}= -{\k \over 2\pi}\int d\Omega ~{\rm Tr} \left[ (-it^a) \left\{
(a_+ - A_+) Q'^\nu
+(Q' \leftrightarrow Q)\right\} \right] \eqno(74)
$$
where $a_+$ is defined by
$$
\partial_- a_+ - \partial_+ A_- +[ A_- ,a_+ ]=0\eqno(75)
$$
The retarded condition is to be used in solving Eq.(75).

It is easy to check that the imaginary part of the
current given by these equations gives the Landau damping effects.
To linear order in the field $A_\mu^a$, we have $J_\mu^a =
\int d^4y~ \Pi_{\mu\nu}^{ab} (x,y) A^{\nu a}(y) $, with
$$
\eqalign{
\Pi_{\mu\nu}^{ab}(x,y) & = \delta^{ab} \int {{d^4k}\over {(2\pi)^4}}~
e^{ik(x-y)}~\Pi_{\mu\nu}(k)\cr
\Pi_{\mu\nu} (k)&= -{\k \over 2\pi} \left[ 4\pi g_{\mu 0}g_{\nu 0}
-k_0 \int d\Omega ~{{Q_\mu Q_\nu }\over {k\cdot Q -\iep}} \right]\cr
}\eqno(76)
$$
The imaginary part
exists only for spacelike momenta and is evidently given by
$$\eqalign{
{\rm Im}~\Pi_{\mu\nu}(k) ~&= k^0 ~
{\k \over 2\pi}  P_{\mu\nu}
\cr
P_{\mu\nu}& \equiv \int d\Omega~ Q_\mu Q_\nu \delta ( k\cdot Q)\cr
&= -k^2 \Theta (-k^2)  {{6\pi }\over {\vert \vk \vert^2}} \left[
{\textstyle 1\over 3} \left(g_{\mu\nu} - {{k_\mu k_\nu }\over k^2}
\right) +
\12 {\tilde P}_{\mu\nu} \right]\cr
{\tilde P}_{0\nu}&={\tilde P}_{\mu 0}=0\cr
{\tilde P}_{ij}&= \delta_{ij} -
{{k_i k_j }\over {\vert \vk \vert ^2}} \cr
}\eqno(77)
$$

As I said before, to the accuracy we are interested in here, the
distinction between
the interacting and incoming field is irrelevant.
One can, of course, go beyond this and include higher order
corrections, using the time-contour approach due to
Schwinger and Bakshi and Mahanthappa, which was rediscovered
by Keldysh $^{19}$. The operator approach is conceptually clearer in indicating
why the retarded condition is appropriate, but a functional
rewriting is useful in going to higher orders.
We define
$$
Z[\eta ]= {{Tr \rho ~{\cal T}_C \exp (iS_{int}+iA\cdot \eta )}\over
Tr \rho}\eqno(78)
$$
where the time-integral goes from $-\infty$ to $\infty$, folds back and
goes from $\infty$ to $-\infty$; ${\cal T}_C$ denotes
ordering along this time-contour. $\rho$ is the thermal
density matrix and $\eta^\mu$ is a source function.
 One can represent $Z[\eta ]$ as a functional integral
$$
Z[\eta ] =\int d\mu (A,c, {\bar c}) \exp (iS_C (A,c, {\bar c}) +
iA_\mu \eta^\mu)
\eqno(79)
$$
The action is again defined on the time-contour. The Green's
functions which arise in perturbatively integrating out the
fields are to be taken as the real-time thermal Green's functions,
as in Eqs.(4,9).
In this time-contour version, the field is given by
is
$$
<A_\mu >= {-i\over Z}{\delta Z\over \delta \eta^\mu (x)}\eqno(80)
$$
We differentiate with respect to $\eta$ on the first branch for the
time-contour. It is evident that we can now define a generator for the
one-particle-irreducible
graphs $\Gamma_C[A]$, such that
${\delta \Gamma_C\over \delta A} = -\eta $. The evolution of a field
configuration in the medium is given by
${\delta \Gamma_C \over \delta A} =0$; we can now set the source to
zero. Separating out the term $\int (\partial_\nu  A_\mu)^2$, which gives
$\partial^2 A_\mu$, we get the current as
$$
<J^a_\mu>= {\delta \Gamma^*_C\over \delta A^a_\mu}\eqno(81)
$$
$\Gamma_C= \int_C {1\over 2}(\partial_\nu  A_\mu )^2+ \Gamma^*_C$.
Eq.(81) is
the same as the expectation value of our operator definition, with
one-particle reducible terms summed up and expressed in terms of
$<A^a_\mu>$ rather than $A^{a in}_{\mu }$.
In writing out Eq.(81), we encounter $A$'s on the second
branch of time. From the operator definition, we see that fields on
the second branch are the conjugates of fields on the first branch.
(In the usual time-contour approach, one considers equations for the
two-point functions, here we consider equations for the fields;
this is the only difference.)

To summarize, the functional transcription of the operator analysis
gives the following. The evolution of fields in the medium is given by
${\delta \Gamma_C^* \over \delta A}=0$, with fields on the second branch
being conjugates of fields on the first branch. One can check that this
gives the retarded prescription we use.
For example, upto linear order in the field $A_\mu$, we get,
$$ \eqalign{
\partial^2 A_\mu &= -i\int_C d^4y~<T_C j_\mu (x)j_\nu (y) >
A^\nu (y)\cr
&= -i\left ( \int_{-\infty}^\infty <T(j_\mu (x)j_\nu (y))> +
\int_{\infty}^{- \infty} <j_\nu (y) j_\mu (x)> \right) A^\nu (y)\cr
&= -i\int \theta (x^0-y^0) <[j_\mu (x), j_\nu (y)]>A^\nu (y)\cr
},\eqno(82)
$$
which agrees with Eq.(71).
\vskip .6cm
\noindent{\twelvebf 8. Derivation of $\G [A]$ from the Boltzmann equation}
\vskip .2in
The equations of motion we obtained
and the expression for the induced current have also been
derived using suitable truncations of the Schwinger-Dyson
equations $^{20}$,
from effective actions for composite operators $^{21}$ and from kinetic
theory $^{22}$. These are all related approaches; here I shall briefly indicate
the kinetic theory approach.
The classical equations of motion for non-Abelian particles,
the so-called Wong equations $^{23}$, are
$$\eqalignno{
m{dx^\mu \over d\tau}&=
p^\mu &(83a)\cr
m{dp^\mu \over d\tau}&= gq^a F_a^{\mu\nu}p_\nu &(83b)\cr
m{dq^a\over d\tau}&= -g f^{abc} (p^\mu A^b_\mu )q^c &(83c)\cr
}
$$
Here $q^a$ represents the classical color charge of the particle.
(The color degrees of freedom can also be described in a phase
space way; the appropriate space is the Lie group modulo the maximal
torus $^{24}$.)

For particles obeying the Wong equations,
the collisionless Boltzmann equation for the distribution function
$f(x,p,q)$ is given by $^{25}$
$$
p^\mu \left[ {\partial \over \partial x^\mu} -g q_a F^a_{\mu\nu}
{\partial \over \partial p_\nu} -g f_{abc}A^b_\mu q^c
{\partial \over \partial q_a}\right] f(x,p,q)=0 \eqno(84)
$$
The Boltzmann equation is invariant under gauge transformations
$gA_\mu \rightarrow  U gA_\mu U^{-1} - U\partial_\mu U^{-1},~
f(x,p,q)\rightarrow f(x,p, U q U^{-1})$. We then seek a perturbative
solution of the form $f= f^{(0)}+g f^{(1)}+...$, where
$f^{(0)}=n_p$ is the equilibrium distribution, appropriately chosen for
bosons and fermions.
The Boltzmann equation, to the first order, gives
$$
p^\mu\left[ {\partial \over \partial x^\mu}
-g f^{abc}A^b_\mu q_c {\partial\over \partial q^a}\right] f^{(1)}=
p^\mu q_a F^a_{\mu\nu} {\partial \over \partial p_\nu} f^{(0)}
\eqno(85)
$$
The color current, to leading order, is
$$
J^a_\mu =g^2 \int [dq]~p_\mu q^a f^{(1)} \eqno(86)
$$
$[dq]$ is a measure for integration over the color charges
which can be constructed in terms of the group coordinates
modulo the maximal torus. By virtue of Eq.(85), $J^a_\mu$ satisfies
the equation
$$
(p\cdot D J^\mu )^a= g^2 p^\mu p^\nu F^b_{\nu\alpha} {\partial \over
\partial p_\alpha} \left( \int dq ~q^aq_b f^{(0)}\right) \eqno(87)
$$
We now integrate this equation over
the magnitude of $\vp $. Defining
$$\eqalign{
J^\mu &= \int d\Omega~ {\cal J}^\mu \cr
{\cal J}^\mu (x,Q)&= \int {{d\vert \vp\vert dp_0 }\over {(2\pi)^3}}
{}~2 \Theta(p_0) \delta (p^2) \vert \vp \vert^2 ~J^\mu \cr
&\equiv {\delta W \over {\delta A_\mu}} -{\k\over 2\pi} Q^\mu A_0, \cr
} \eqno(88)
$$
we see that Eq.(87) for the current becomes the zero-curvature
condition Eq.(75); the rest of the analysis is as
before, leading to Eqs.(72,73). This derivation from kinetic theory
shows that the hard thermal loops are classical; they include thermal
fluctuations, not quantum fluctuations. Of course, we have already
seen this from another point of view, viz., the thermal part of the
propagator is on-shell and so, the $T$-dependent part of a
one-loop diagram describes the tree-level absorption and emission
of particles from the heat bath.
\vskip .6cm
\noindent{\twelvebf 9. An auxiliary field and Hamiltonian analysis}
\vskip .2in
We now turn to the question of how we can use the effective action
$\G [A]$.
There are two related but different contexts in which we need
the expression for $\Gamma [A]$. The first is in
setting up thermal perturbation theory. The version of $\Gamma [A]$
as given in Eqs.(64,68) is probably best suited for this purpose.
The resummed perturbation theory can be set up as follows.
We introduce a splitting
of the Yang-Mills action as
$$\eqalignno{S&=S_0-c\Gamma\cr
	     S_0&=\int d^4x~\left( ~{\textstyle-{1\over 4}}F^2 \right)
{}~+\Gamma&(89)\cr}
$$
\noindent We define propagators and vertices and start off the
perturbative expansion using $S_0$. $c\Gamma$ will be
treated
as a `counterterm', nominally one order higher in the thermal loops
than $S_0$. Eventually of course $c$ is taken to be 1, so that
we are only achieving a rearrangement of terms in the
perturbative expansion. (As usual we must have gauge fixing and
ghost terms.) Using this procedure one can
calculate quantities which require resummations such as the gluon
decay rate in the plasma $^9$. In Eq.(89) we have not displayed  the
quark terms. In addition to the quark kinetic energy terms, there
are hard thermal loops which give a $T$-dependent mass to the
quarks $^{12,26}$. We have not discussed this so far, since
Chern-Simons theory does not give any new insights on this
question. This mass term is actually given by
$$\eqalign{
\G ^{(q)}[q,A] &= \int d^4x d^4y~ {\bar q}(x) \Sigma (x,y) q(y)\cr
\Sigma (x,y) &={g^2T^2 C_F\over 64 \pi}\int d\Omega ~\gamma \cdot Q
{}~ F(x,y)\cr
(i Q\cdot D )F(x,y)&= \delta^{(4)}(x-y) \cr }
\eqno(90)
$$
where $C_F= t^at^a$ is the value of the
quadratic Casimir for the quark representation.
This action must also be added to and subtracted from the QCD
action and treated in a manner
analogous to $\G [A]$ to set up the perturbation theory properly.

In addition to its role in rearrangement of perturbation theory,
we can
also use $\Gamma [A]$ added to the usual Yang-Mills action, viz.,
$$
S_{eff}= \int d^4x~ \left( -{\textstyle {1\over 4}} F^a_{\mu\nu}
F^{\mu\nu}_a \right)~+ \G [A],\eqno(91)
$$
as an
effective action for the soft modes. (This is the spirit of
Eqs.(72,73).)
The nonlocality of $\Gamma [A]$ makes it somewhat difficult to
handle in this context. It is useful to rewrite $\Gamma [A]$ using
an auxiliary field which makes the equations of motion
local $^{3}$. The auxiliary field is also useful in setting up
a Hamitonian analysis of the dynamics of the soft modes.
The auxiliary field we
use will be an $SU(N)$-matrix field $G(x,{\vec Q})$ which is a function
of $x$ and ${\vec Q}$, i.e., defined on ${\cal M}^4 \times S^2$,
${\cal M}^4$ being
Minkowski space. Further $G(x,{\vec Q})$ must satisfy the condition
$G^{\dagger}(x, {\vec Q})= G(x, -{\vec Q})$. The action is given by
$$\eqalignno{
{\cal S}= \int -{\textstyle{1\over 4}}F^2~+~& \k \int d\Omega~\Biggl[
d^2x^T~ {\cal S}_{WZNW}(G) ~+~{1\over \pi}\int d^4x~ {\rm Tr}(
G^{-1}\partial_- G~A_+ \cr
&~-~ A_- \partial_+G~G^{-1}+A_+ G^{-1}A_- G -
A_+ A_-) \Biggr] &(92)\cr}
$$
where ${\cal S}_{WZNW}(G)$ is the WZNW action of Eq.(58)
with $U$ replaced by $G$.
The quantity in the square brackets in Eq.(92) is the gauged WZNW
action $^{27}$. It is invariant under gauge transformations with $G$
transforming as $G\rightarrow G'= h(x) G~ h^{-1}(x),~h(x)\in SU(N)$.
The equations of motion for the action (92) are
$$
\partial_+ A_- - \partial_- a_+ ~+~ [a_+ ,A_- ]=0 \eqno(93a)
$$
$$
a_+ \equiv G A_+ G^{-1} - \partial_+ G~G^{-1}\eqno(93b)
$$
$$
(D_\mu F^{\mu\nu})^a -J^{\nu a}=0\eqno(94a)
$$
$$\eqalignno{
J^{\nu a}&= -{\k \over 2\pi } \int d\Omega~ {\rm Tr}[ \{ (-it^a ) (a_+ -A_+)
Q'^\nu \} +(Q'\leftrightarrow Q)] &(94b)\cr
&= -{\k \over 2\pi } \int d\Omega~ {\rm Tr}[  (-it^a) \{ G^{-1}
D_- G~Q^\nu ~-~ D_+G ~G^{-1} Q'^\nu \}] &(94c)\cr}
$$
Clearly these equations are equivalent to Eqs.(72,74,75); the only
difference is that the equation defining $a_+$ in Eq.(74), viz.
Eq.(75), is now
obtained as the equation of motion (93a) for $G$. Notice that the
current in Eq.(94c) looks like the current of a matter field; thus
except for the fact that $G$ depends on $\vec Q$, QCD,
with hard thermal
loops added, is no stranger than Yang-Mills theory coupled to a
matter field. The equations of motion for $G$ has no independent
solutions, so that we are not changing the physical degrees of freedom
by introducing the auxiliary field. This can be seen as follows.
We can parametrize the potentials $A_{\pm}$ in terms an
$SU(N)$-matrix $V(x,\vQ)$ as
$$
A_{+}=~ -\partial_{+} V~V^{-1},~~~~~~~~A_{-}=~ -\partial_{-}V'~V'^{-1}
\eqno(95)
$$
where $V' (x, \vQ)= V(x,-\vQ)$. The general solution to Eq.(93) is
then given by
$$
G(x,\vQ)~=~ V(x, -\vQ) B(x^{+},x^T, \vQ) C(x^-, x^T, \vQ) V^{-1}(x,\vQ
)\eqno(96)
$$
where  $C$ is an arbitrary $SU(N)$-matrix depending on the variables indicated
and $B$ is given by $C$ with $\vQ \rightarrow -\vQ$. The matrices
$B,C$
represent the new or independent degrees of freedom for the field $G$.
Notice, however, that the parametrization (95) of the potentials has
redundant variables; a transformation $V\rightarrow V U (x^-, x^T,\vQ) $,
with a corresponding change in $V'$, leaves the potentials invariant.
Further, ordinary gauge transformations act on the matrix $V$ as $V
\rightarrow h(x)V$. Thus the physical subspace for the components
$A_{\pm}$ is given in terms of the matrices $V$ with the identifications
$$
V(x,\vQ) \sim h(x) V(x,\vQ) U(x^-,x^T,\vQ) \eqno(97)
$$
The gauge freedom of multiplying $V$'s by matrices which do not depend
on $x^+$, viz., $U$'s in Eq.(97), shows that the we can reduce $G$ to just
$V(x,\vQ) V^{-1}(x,\vQ)$. There are thus no new real dynamical degrees
of freedom in $G$. The action can be simplified for this $G$ to
$$
\Gamma [A,G]=~ - \k \int d\Omega ~d^2x^T ~S_{WZNW}\bigl( V^{-1}(x,-\vQ )
V(x,\vQ )
\bigr) \eqno(98)
$$
This agrees with the expression obtained by Feynman graph evaluation of
the hard thermal loops.

The Hamiltonian for the effective action (92) can be obtained by
following the usual analysis for the WZNW-action; it is given by $^3$
$$
{\cal H}=~\int d^3x~\left\{ {{E^2+B^2}\over 2} ~+~{\k \over 8\pi} \int
d\Omega~ {\rm Tr}\left[  (D_0G~ D_0G^{-1})+(\vQ \cdot {\vec D}G~\vQ
\cdot {\vec D}G^{-1} \right] -A^a_0 {\cal G}^a \right\} \eqno(99)
$$
where $F^a_{0i}=E^a_i,~ F^a_{ij}=\epsilon_{ijk}B_k^a$ and
$$
{\cal G}^a= ~({\vec D}\cdot {\vec E})^a ~+~ {\k \over 2\pi}\int d\Omega~
{\rm Tr}\bigl[ (-it^a) (G^{-1} D_- G~-~ D_+ G~G^{-1})\bigr]
\eqno(100)
$$

${\cal G}^a=0$ is the Gauss law of the theory; it is also the time
component of the equation of motion for the gauge field.
Expression (99)
makes it clear that the Hamiltonian is positive for all configurations
which are physical, i.e., obey the Gauss law. (It may be worth recalling
that, even for the Maxwell theory, the canonical Hamiltonian is positive
only for physical configurations.)  We find, comfortingly, that
 the effective
theory for the soft modes has positive energy.

The Hamiltonian analysis is most easily carried out, as in the usual
cases, in the gauge $A_0^a=0$. We start by defining the currents
$$
J_+ ~=~ {\k \over 4\pi} D_+ G~G^{-1}~=~ (-it^a)J^a_+
$$
$$
J_-~=~ -{\k \over 4\pi} G^{-1}D_-G~=~ (-it^a)J^a_- \eqno(101)
$$
By virtue of the property $G^{-1}(x,\vQ )= G(x, -\vQ )$, these are related
by $J_+ (x, -\vQ )= J_- (x, \vQ )$.  The Hamiltonian can be written in
terms of these currents as
$$
{\cal H}=~ \int d^3x~\left\{ {{E^2+B^2}\over 2} ~+{2\pi \over \k} \int
d\Omega~ (J^a_+ J^a_+ ~+~ J^a_- J^a_- ) \right\} \eqno(102)
$$
We have chosen the $A_0^a=0$ gauge; Gauss law must henceforth be
imposed as a constraint. ( For fixed $\vQ$, the integrand of the
second term involving the square of $J_+$ and the square of
$J_-$ is the Sugawara form of the Hamiltonian, well-known
in the context of two-dimensional
current algebras.) The equations of motion in the $A_0^a=0$ gauge
are
$$
\eqalignno{
E^a_i~ &=\partial_0 A^a_i &(103a)\cr
\partial_0 E^a_i ~+~ \epsilon_{ijk} (D_j B_k)^a ~&=~ \int d\Omega ~Q_i
(J^a_+ ~-~J^a_- ) &(103b)\cr
(D_- J_+)^a ~&=~ -{\k \over 8\pi } E^a_i Q_i &(103c)\cr }
$$
Equation (103a) is just the definition of the electric field; however, in a
Hamiltonian approach, it is an equation of motion and we have displayed
it as such.

The commutation rules must be such that Eqs.(103)
follow as the Heisenberg equations of motion for the Hamiltonian
(102). Knowing the current algebra of the WZNW-theory, we can make a
guess as to what the appropriate commutation rules are for our
problem and verify them by checking
that they lead to Eqs.(103) starting from the Hamiltonian (102).
The commutation rules are then seen to be
$$
\eqalignno{
[~E^a_i (\vx ), A^b_j(\vx ' ) ~] ~&= -i\delta^{ab}\delta_{ij} \delta(\vx -\vx
')
&(104a)\cr
[~E^a_i(\vx ), J^b_{\pm} (\vx ')~]~&= {\pm}i {\k \over 4\pi} Q_i \delta^{ab}
\delta(\vx -\vx') &(104b)\cr
[~J^a_{\pm} (\vx, \vQ), J^b_{\pm}(\vx' , \vQ' )~]~&= if^{abc}
J^c_{\pm}\delta(\vx -\vx') \delta(\vQ, \vQ')\cr
& ~~~~~{\mp}{\k \over 4\pi}
Q_i(D_x)^{ab}_i \delta(\vx -\vx') \delta (\vQ, \vQ') &(104c)\cr
[~J^a_+ (\vx, \vQ), J^b_- (\vx', \vQ')~]~&=0 &(104d)\cr}
$$
All other commutators vanish. $\delta (\vQ, \vQ')$ stands for the
$\delta$-function on the sphere
corresponding to the unit vector $\vQ$, i.e., $\int d\Omega_{Q'}
{}~\delta (\vQ,\vQ') f(\vQ') =f(\vQ)$.

The algebra (104) is an interesting extension of the
usual WZNW-current algebra; we are in four dimensions and have a
gauge field as well. We can check that the commutation rules (104) obey the
Jacobi identity. Since we are postulating the algebra, this is a necessary
check. Of course, commutation rules can also be obtained from
the action by standard quantization procedures $^3$.
The condition $J^a_+ (x -\vQ)~-~J^a_- (x,\vQ)=0$
has to be imposed as a constraint, just like the Gauss law.
\vskip .6cm
\noindent{\twelvebf 10. Plasma waves}
\vskip .2in
Long wavelength and low frequency plasma waves are the classical solutions
of the effective theory Eq.(91). It also includes effects such as the screening
of Coulomb fields. These features can be seen by examining the Abelian
case or electrodynamics.
In this case, the terms in $\Gamma$ which are cubic or higher
order in $A_\mu$ are zero and in terms of the Fourier components of $A_\mu$,
we can write
$$
S_{eff}={\textstyle{1\over2}}\int {d^4k\over(2\pi)^4} ~
A_\mu(-k)M^{\mu\nu}(k) A_\nu(k)
\eqno(105a)
$$
\noindent where
$$
M^{\mu\nu}=(-k^2g^{\mu\nu}+k^\mu k^\nu)+
{\k \over 2\pi } \left[4\pi g^{\mu 0} g^{\nu 0}
-k^0 \int d\Omega ~{{Q^\mu Q^\nu} \over k\cdot Q}\right].
\eqno(105b)
$$
$k^\mu M_{\mu\nu}=0$ in accordance with the requirement of
gauge invariance.
$\k = N_F e^2T^2 /6$ for electrodynamics;
we have restored the coupling constant $e$ in this expression.
For the plasma waves, we will see that the wavevector
is timelike; hence the imaginary part of $M_{\mu\nu}$ is irrelevant.

We can split $A_\mu$ into a gauge dependent part and gauge
invariant components as
$$A_\mu =k_\mu \Lambda (k) +\alpha_\mu +\beta_\mu\eqno(106)
$$
\noindent where $\Lambda$ shifts under gauge transformations and
$\alpha_\mu$, $\beta_\mu$ are gauge invariant. We take
$$
\alpha_0=({{\vec k}^2\over {\vec k}^2-k_0^2})\phi ~~~~
\alpha_i={k_0\over\sqrt{{\vec k}^2}}e^{(3)}_i~({{\vec k}^2\over
{\vec k}^2-k_0^2})\phi
$$
$$\beta_0=0~~~~
\beta_i=e_i^{(\lambda)}a_\lambda, ~~~~\lambda=1,2.\eqno(107)
$$
(Here we are considering fields off-shell and so
$k^0 \neq \vert \vec k
\vert$.) The $e_i$'s form a triad of spatial unit vectors which
may be taken as
$$e_i^{(3)}={k_i\over\sqrt{{\vec k}^2}},~~~~i=1,2,3,$$
$$e^{(1)}=(\epsilon_{ij}{k_j\over\sqrt{k_T^2}},0),~
{}~~~e^{(2)}=({k_3k_i\over\sqrt{k_T^2{\vec k}^2}},-\sqrt{k_T^2\over
{\vec k}^2}), ~~~~i=1,2, \eqno(108)
$$
\noindent where $k_T^2=k_1^2+k_2^2$. Notice that $k_ie_i^{(\lambda)}=0$,
$\lambda=1,2$. $\phi$ and $a_\lambda$ are the gauge invariant degrees
of freedom in Eq.(106). When the mode
decomposition (106) is used in Eq.(105) we get
$$
S_{eff}={\textstyle{1\over2}}\int {d^4k\over (2\pi)^4}
{}~\left[~\beta_i (-k)({{\vec k}^2\delta_{ij}-k_ik_j\over{\vec k}^2})
M^T(k)\beta_j(k)+\phi(-k)M^L(k)\phi(k)\right]
+\int\phi J^0+\beta_i J^i\eqno(109)
$$
\noindent where
$$\eqalign{
M^T(k)&=k_0^2-{\vec k}^2- \Omega_T^2 (k_0, \vk )\cr
\Omega_T^2&= \k \left[~{k_0^2\over{\vec k}^2}
+(1-{k_0^2\over{\vec k}^2}){k_0\over 2|{\vec k}|}L~\right]\cr}
$$
$$\eqalign{
M^L(k)&={{\vec k}^2\over (k_0^2-\vk ^2)}
\left( k_0^2 -\vk^2 - \Omega_L^2 \right)\cr
\Omega_L^2&= 2\k ~\left( {k_0^2-\vk^2}\over \vk^2\right)
({k_0\over2|{\vec k}|}L ~-1)\cr}
\eqno(110)
$$
$$
L=\log({k_0+ |\vk |\over k_0-|\vk |}).\eqno(111)
$$

\noindent
We have also included an interaction term with a conserved source
$J_\mu$ in Eq.(109); i.e., we include $\int A_\mu J^\mu$ and simplify it
using Eq.(106). From Eq.(108) we see that the interaction between
charges in the plasma is governed by $(M^L)^{-1}$ which shows the
Debye screening with a Debye mass $m_D=\sqrt{2\k }$.
The action (109) can also give free wavelike solutions.
The dispersion rules for these plasma waves would
be $M^T=0$ for the transverse waves and
$({{\vec k ^2-k_0^2}\over {\vec k
^2}})M^L=0$ for the
longitudinal waves $^{28}$. (The extra factor multiplying $M_L$ is from
rewriting $\phi(k)$ in terms of the potential.)

Non-Abelian plasma waves can be defined in a similar way as propagating
solutions to the equations of motion given by Eq.(91). For plasma waves in
Abelian subgroups, only terms in $\G$ upto the quadratic order
in $A_\mu$ are important; the analysis is the  same as for
the Abelian plasma waves with
$m_D^2= 2\k =(N+{1\over 2}N_F){g^2T^2\over3}$ and
$\k =(N+{1\over 2}N_F){g^2T^2\over6}$
in $M^T(k)$.

It is interesting to consider these waves also from the Hamiltonian point
of view. We can choose $A_0=0$ gauge and introduce the parametrization
$$\eqalignno{
A_i &= e^{(3)}_i a + e^{(\lambda )}_i a_\lambda & (112a)\cr
E_i&= e^{(3)}_i \Pi +e^{(\lambda )}_i \Pi_\lambda &(112b)\cr }
$$
We have dropped the color indices, since we are in an Abelian subgroup.
The Hamiltonian becomes
$$
{\cal H}= \int d^3x~ \12 \left[ \Pi^2 +\Pi_\lambda ^2 +
\partial_ja_\lambda \partial_ja_\lambda +{4\pi \over \k }
\int d\Omega (J_+^2+J_-^2)\right] \eqno(113)
$$
The commutation rules are
$$
\eqalignno{
[\Pi (\vx ), a(\vy )]= -i\delta (\vx -\vy ),~~~~~~~~
&[\Pi_\lambda (\vx ),
a_\lambda' (\vy )]= -i\delta_{\lambda \lambda'} \delta (\vx -\vy )
&(114a)\cr
[\Pi (\vx ), J_{\pm}(\vy )] = \pm i{\k \over 4\pi} \vQ
\cdot e^{(3)}
\delta (\vx -\vy ), ~~~~~~~~&[\Pi_\lambda (\vx ), J_{\pm}(\vy )]=
\pm i{\k\over 4\pi} \vQ \cdot e^{(\lambda )} \delta (\vx -\vy )
&(114b)\cr}
$$

The creation operators for the eigenstates obey the
eigenvalue equation
$$
[{\cal H}, \alpha ]=\omega \alpha \eqno(115)
$$
For the longitudinal modes, we can take
$$
\alpha = \int d^3x~ e^{-i\vk \cdot \vx}
\left[ c_1 \Pi +c_2 a +\int d\Omega (c_3 J_+ +c_4 J_-) \right]
\eqno(116)
$$
Using this in the eigenvalue equation, we find
$$\eqalignno{
\omega ~c_1 &=i{\k \over 4\pi} \int d\Omega ~ \vQ\cdot e^{(3)}
(c_3 -c_4)&(117a)\cr
c_3 &= -{i\over {\vert \vk \vert }} {{\vk \cdot \vQ}\over
(\omega+\vk \cdot \vQ )} ~c_1&(117b)\cr}
$$
and $c_2=0,~c_4=c_3(-\vQ )$. Using Eq.(117b) in Eq.(117a), we find that a
nonzero
solution requires
$$
\omega - {\k\over 2\pi} \int d\Omega ~ {{(\vk \cdot \vQ )^2}\over
{\vert \vk \vert^2 (\omega - \vk \cdot \vQ )}} =0 \eqno(118)
$$
This gives the $(\omega ,\vk )$-relation for the longitudinal modes.
The creation operator is then
$$
\alpha_L  (\vk )=
\Pi (\vk ) -{i \over {\vert \vk \vert^2}} \int d\Omega ~\left[
{{\vk \cdot \vQ}\over {(\omega +\vk \cdot \vQ )}} J_+ (\vk ) -
{{\vk \cdot \vQ}\over {(\omega -\vk \cdot \vQ )}}J_-(\vk ) \right]
\eqno(119)
$$
A similar analysis can be done for the transverse modes, the creation
operator being
$$
\alpha_T (\vk )= \Pi_\lambda (\vk ) +i {{\vert \vk \vert^2}\over
\omega} a_\lambda (\vk ) -i \int d\Omega~ \left[
{\vQ \cdot e^{(\lambda )}\over (\omega +\vk \cdot \vQ )}J_+ (\vk )-
{\vQ \cdot e^{(\lambda )}\over (\omega -\vk \cdot \vQ )}J_- (\vk )
 \right] \eqno(120)
$$

So far we have considered plasma waves in Abelian subgroups.
One can also
construct more general solutions $^{29}$. Consider fields of
the form $ A^a_\mu = A^a_\mu (p\cdot x)$, i.e., the field depends only
on the combination $s= p\cdot x$ where $p^\mu$ is a timelike vector.
The zero-curvature condition (75) can be solved as
$$
a_+ = \12 {\dot A}\cdot Q'~ {p\cdot Q \over p\cdot Q'} \eqno(121)
$$
where ${\dot A}= {dA\over ds}$. Writing $A_\mu = \epsilon_\mu (p)
h_a (s) (-it^a )$,
we find the equations of motion,
$$
\eqalignno{
p^\mu p_\mu {\ddot h_3} ~+\Omega_L^2 h_3 &=0 &(122a)\cr
p^\mu p_\mu {\ddot h_1}~+ \Omega_T^2 h_1 +(h_2^2 +h_3^2 )h_1&=0
&(122b)
\cr}
$$
(We have taken the gauge group to be $SU(2)$ for simplicity;
there are also equations with cyclic permutations of the labels.)
Define
$$
k^\mu = {p^\mu \Omega  \over {\sqrt { p^\alpha
p_\alpha}}},~~~~~~~~~k^2= \Omega^2 , \eqno(123)
$$
for transverse and longitudinal cases.
The longitudinal mode has a solution of the form $e^{ik\cdot x}$;
the relation between $k^0$ and $\vk$ is given by
$k^\mu k_\mu =\Omega_L^2$, which is the same relation as for the
Abelian waves. For the transverse modes, the $(k^0,\vk )$-relation is
again the same as for the Abelian case; the amplitudes depend on
$\tau =k\cdot x$ and obey the
nonlinear equations
$$
{\ddot f_a}+ [ 1+(\epsilon_{ab}f_b)^2 ]f_a =0,~~~~~~~~a,b=1,2
\eqno(124)
$$
where now ${\dot f}= {df \over d\tau}$.
These have been further studied in ref.[29].
\vskip .6cm
\noindent{\twelvebf 11. Magnetic screening and magnetic mass}
\vskip .2in
Hard thermal loops and the effective action are important in
setting up the resummed perturbation theory. Now, $\G [A]$ is also
an electric mass term for gluons and
properly incorporating $\Gamma [A]$ in any calculation eliminates some of the
infrared singularities. However, there would still remain some
singularities since the
static magnetic interactions are not screened. It is possible to
introduce an infrared cutoff by hand to screen the magnetic
interactions and make perturbative calculations well defined.
Interestingly, a gauge invariant magnetic screening term can also be
constructed using
Chern-Simons related techniques $^{30}$.
Of course, in principle, one should not have to introduce
such a cutoff by hand.
It is generally believed that
for QCD at high temperatures there is a magnetic mass term
which screens
the static magnetic interactions (or more generally magnetic fields with
spacelike
momenta). One way to understand how this might happen is
as follows $^{31}$. In the
standard imaginary-time formalism for
equilibrium statistical mechanics, bosonic fields are periodic
in the imaginary time $\tau$ with period $1/T$,
i.e., $\phi(\vx, \tau + 1/T )=\phi(\vx ,\tau )$.
The propagator has the form
$$
G(x,y)= T\sum_n\int {{d^3p}\over {(2\pi )^3}} {e^{-ip(x-y)}
\over (\omega_n^2 +|\vp |^2 )} \eqno(125)
$$
$\omega_n= 2\pi n T,~n=0, \pm 1, \pm 2,...,$ are the Matsubara frequencies.
At high temperatures, because of the structure of the denominator,
we expect only $T=0$  mode to be important. There is
no $\tau$-dependence for this mode, and thus we expect QCD to
reduce to
three-dimensional QCD. The coupling constant of this
reduced theory is ${\sqrt{g^2 T}}$. For QCD in three dimensions
we expect a mass gap ($\sim g^2T$) and this is effectively
the magnetic mass of the high temperature four-dimensional QCD.
The magnetic mass term is expected to be gauge invariant and
parity even, although not necessarily local.
It must be Lorentz invariant if we include the overall motion of
the plasma. Also we expect such a term to be relevant only for the
spatial components of the gauge potential. A term which has all
these properties is given by
$$
{\tilde \G [A]} = - M^2 \int d\Omega~ K[A_n, A_{\bar n}] \eqno(126)
$$
where $A_n= \12 A_in_i,~ A_{\bar n}= \12 A_i{\bar n}_i$ and
$$
n_i= (-\cos \theta \cos \varphi -i \sin \varphi,~-\cos \theta
 \sin \varphi +i \cos \varphi,~ \sin \theta~) \eqno(127)
$$
(This is in the rest frame of the plasma.)
Notice that $n_i$ is a complex
three-dimensional null vector; it takes
over the role of the null vector $Q^\mu$ of the electric mass term.
${\tilde \G}$ involves,
in the rest frame, only the
spatial components of $A_\mu$ as expected for a magnetic mass term.
$I(A_n),~I(A_{\bar n})$ are defined using $n\cdot x$ and ${\bar n}
 \cdot x$
in place of $z,{\bar z}$ in Eq.(64). Thus $K$ in Eq.(126) is
$K[\12 A\cdot Q, \12 A\cdot Q' ]$ of Eq.(68) with
$Q^\mu\rightarrow (0,n_i),~Q'^\mu \rightarrow
(0,{\bar n}_i )$.

Consider the simplification of the quadratic term in Eq.(126). Using
$$\eqalign{
\int d\Omega~ n_i {\bar n}_j &= {8\pi \over 3} \delta_{ij}\cr
\int d\Omega ~{k\cdot {\bar n} \over {k\cdot n}} n_i n_j&=
{8\pi \over 3}
\left[ {k_ik_j \over \vk^2}-{1\over 2} \left(\delta_{ij}-
{k_ik_j \over \vk^2}\right)\right] \cr}\eqno(128)
$$
we find
$$
{\tilde \G} = -{M^2 \over 2}\int {d^4k \over (2\pi )^4}~
A^a_i (-k) \left( \delta_{ij}-
{k_ik_j\over \vk^2}\right) A^a_j(k) ~+ {\cal O}(A^3) \eqno(129)
$$
Thus ${\tilde \G}$ does give screening of transverse magnetic
interactions, with a screening mass $M$.

The higher order
terms in ${\tilde \G}$ can also be evaluated in a
straightforward fashion, noting
that the basic change is replacing $Q^\mu$ by $(0,n_i) $ and
$Q'^\mu $ by $(0,{\bar n}_i )$.

Of course, there can
be other ways of representing magnetic screening.
What is notable about the above way of writing the magnetic
screening effects is its evident kinship to the electric mass term
and the Chern-Simons eikonal.
Actually, in an arbitrary frame, both
$\G$ and ${\tilde \G}$ look the same, except for the
choice of different orbits of the Lorentz group. Of course, we do
not know what $M$ is; no calculation even with summations of
an infinity of diagrams exists at this point. There are some
estimates based on lattice and other approaches to three-dimensional
QCD. Even without a value for $M$, one can use
${\tilde \G}$ as a gauge-invariant infrared cutoff in
perturbative caluclations. One can also try to estimate $M$ by
some self-consistent method. Recall that for the Nambu-Jona Lasinio
model $^{32}$, once we know the spinor structure of the mass term, viz.,
$m{\bar q}q$, we can write
$$
{\cal L}= {\cal L}_{NJL} +m {\bar q}q -m{\bar q}q \equiv
{\cal L}_0 - m {\bar q}q \eqno(130)
$$
We then calculate the one-loop effective action using ${\cal L}_0$,
taking the subtracted term $-m{\bar q}q$ formally as a one-loop
term. The vanishing of the correction to the mass then becomes
the gap equation determining $m$. A similar calculation can be
attempted here.
The trouble is that one cannot really stop with the one-loop term;
contributions from all loops are generally significant.
It is possible that the self-consistency equations will simplify
in some cases,
such as perhaps the planar diagram approximation.
\vglue 0.6cm
\leftline{\twelvebf 12. Acknowledgements}
\vglue 0.4cm
Finally I
thank Professor J.E. Kim and others
involved in organizing the Mt. Sorak summer school for inviting me
to give these lectures. The summer school was an occasion,
both pleasant and intellectually stimulating, and it gave me the
opportunity to get to know a great country and a nice people.

This work was supported in part by
NSF grant PHY-9322591.
\vglue 0.6cm
\leftline{\twelvebf 13. References}
\vglue 0.4cm
\medskip
\itemitem{1.} R.Efraty and V.P.Nair, {\twelveit Phys.Rev.Lett.}
{\twelvebf 68} (1992) 2891; {\twelveit Phys.Rev.} {\twelvebf D47} (1993)
5601.
\itemitem{2.} R.Jackiw and V.P.Nair, {\twelveit Phys.Rev.}
{\twelvebf D48} (1993) 4991.
\itemitem{3.} V.P.Nair, {\twelveit Phys.Rev.} {\twelvebf D48} (1993) 3432;
{\it ibid.} {\twelvebf D50} (1994) 4201.
\itemitem{4.} B.Svetitsky and L.Yaffe, {\twelveit Nucl.Phys.}
{\twelvebf B210} (1982) 423 and references therein;
R.Pisarski and F.Wilczek, {\twelveit Phys.Rev.} {\twelvebf D29}
(1984) 338.
\itemitem{5.} For a recent review, see  B.Petersson,
{\twelveit Nucl.Phys.
(Proc.Suppl.)} {\twelvebf B30} (1993) 66.
\itemitem{6.} See, for example, C.P.Singh, {\twelveit Phys.Rep.}
{\twelvebf 236} (1993) 147.
\itemitem{7.} Reference 6; K.Rajagopal and F.Wilczek, {\twelveit Nucl.Phys.}
{\twelvebf B379} (1993) 395; {\twelveit ibid.} {\twelvebf B404}
(1993) 577; articles by R.Pisarski and U.Heinz in {\twelveit Banff/CAP Workshop
on Thermal
Field Theory}, F.C.Khanna, {\it et al} (eds.) (World Scientific,
Singapore, 1994).
\itemitem{8.} V.P.Silin, {\twelveit Sov.J.Phys.JETP}
{\twelvebf 11} (1960) 1136;
V.N.Tsytovich, {\twelveit Sov.Phys.JETP}, {\twelvebf 13} (1961) 1249;
O.K.Kalashnikov and V.V.Klimov, {\twelveit Sov.J.Nucl.Phys.},
{\twelvebf 31} (1980) 699;
V.V.Klimov, {\twelveit Sov.J.Nucl.Phys.} {\twelvebf 33} (1981) 934;
{\twelveit Sov.Phys.JETP} {\twelvebf 55} (1982) 199;
H.A.Weldon, {\twelveit Phys.Rev.D} {\twelvebf 26} (1982) 1394.
\itemitem{9.} R.Pisarski, {\twelveit Physica A} {\twelvebf 158} (1989)
246; {\twelveit Phys.Rev.Lett.} {\twelvebf 63} (1989) 1129;
E.Braaten and R.Pisarski, {\twelveit Phys.Rev.} {\twelvebf D42} (1990)
2156; {\twelveit Nucl.Phys.} {\twelvebf B337} (1990) 569; {\twelveit
ibid.} {\twelvebf B339} (1990) 310; {\twelveit Phys.Rev.} {\twelvebf
D45} (1992) 1827.
\itemitem{10.} R.Kobes, G.Kunstatter and A.Rebhan, {\twelveit Nucl.Phys.}
{\twelvebf B355} (1991) 1.
\itemitem{11.} C.Itzykson and J.B.Zuber,
{\twelveit Quantum Field Theory} (McGraw Hill, New York, 1980)
pp.594-599.
\itemitem{12.} J.Frenkel and J.C.Taylor, {\twelveit Nucl.Phys.}
{\twelvebf B334} (1990) 199; J.C.Taylor and S.M.H.Wong, {\twelveit
Nucl.Phys.} {\twelvebf B346} (1990) 115.
\itemitem{13.} R.Jackiw and S.Templeton, {\twelveit Phys.Rev.}
{\twelvebf D23} (1981) 2291; J.Schonfeld, {\twelveit Nucl. \break
Phys.}
{\twelvebf B185} (1981) 157; S.Deser, R.Jackiw and S.Templeton,
{\twelveit Phys.Rev.Lett.} {\twelvebf 48} (1982) 975;
{\twelveit Ann.Phys.} {\twelvebf 140} (1982) 372.
\itemitem{14.} E.Witten, {\twelveit Commun.Math.Phys.}
{\twelvebf 121} (1989) 351.
\itemitem{15.} D.Gonzales and A.N.Redlich,
{\twelveit Ann.Phys.(N.Y.)}
{\twelvebf 169} (1986) 104;
B.M.Zupnik, {\twelveit Phys.Lett.} {\twelvebf B183} (1987) 175;
G.V.Dunne, R.Jackiw
and C.A.Trugenberger, {\twelveit Ann.Phys.(N.Y.)}
{\twelvebf 149} (1989) 197; M.Bos and V.P.Nair,
{\twelveit Phys.Lett.} {\twelvebf B223} (1989) 61; {\twelveit
Int.J.Mod.Phys.} {\twelvebf A5} (1990) 959;
S.Elitzur, G.Moore, \break
A.Schwimmer and
N.Seiberg, {\twelveit Nucl.Phys.} {\twelvebf B326} (1989) 108;
A.P. Polychronakos,
{\twelveit Ann. Phys.} {\twelvebf 203} (1990) 231.
\itemitem{16.} S.P. Novikov, {\twelveit Usp.Mat.Nauk}
{\twelvebf 37} (1982) 3; E. Witten,
{\twelveit Commun.Math.Phys.} {\twelvebf 92} (1984), 455;
V.G.Knizhnik and
A.B.Zamolodchikov, {\twelveit Nucl.Phys.}
{\twelvebf B247} (1984) 83.
\itemitem{17.} A.M.Polyakov and P.B.Wiegmann, {\twelveit Phys.Lett.}
 {\twelvebf 141B} (1984) 223.
\itemitem{18.} See, for example, S.S.Schweber, {\twelveit An
Introduction to Relativistic Quantum Field Theory} (Harper and Row,
New York, 1964) pp-742-764.
\itemitem{19.} J.Schwinger, {\twelveit J.Math.Phys.} {\twelvebf 2}
(1961) 407; P.M.Bakshi and K.Mahanthappa, {\twelveit J.Math.Phys.}
{\twelvebf 4} (1963) 1; {\twelveit ibid.} {\twelvebf 4} (1963)
12; L.V.Keldysh, {\twelveit Sov.Phys. JETP} {\twelvebf 20} (1964)
1018.
\itemitem{20.}  J.P.Blaizot and E.Iancu, {\twelveit Phys.Rev.Lett.}
 {\twelvebf 70} (1993) 3376;
{\twelveit Nucl.Phys.} {\twelvebf B 417} (1994) 608.
\itemitem{21.}  R.Jackiw, Q.Liu
and C.Lucchesi, {\twelveit Phys.Rev.} {\twelvebf D49} (1994) 6787.
\itemitem{22.} P.F.Kelly {\it et al}, {\twelveit Phys.Rev.Lett.}
{\twelvebf 72} (1994) 3461; {\twelveit Phys.Rev.}
{\twelvebf D50} (1994) 4209.
\itemitem{23.} S.Wong, {\twelveit Nuovo Cim.} {\twelvebf 65A}
(1970) 689.
\itemitem{24.}See, for example, A.W.Knapp, {\twelveit Lie Groups,
Lie Algebras and Cohomology} (Princeton University Press, Princeton,
1988); A.Pressley and G.Segal, {\twelveit Loop Groups} (Oxford:
Clarendon Press, 1986); J.Sniatycki, {\twelveit Geometric
Quantization and Quantum Mechanics} (Springer-Verlag, New York,
1980).
\itemitem{25.} U.Heinz, {\twelveit Phys.Rev.Lett.}
{\twelvebf 51} (1983) 351; {\twelveit Ann.Phys. (N.Y.)} {\twelvebf 161}
(1985) 48; {\twelveit Ann.Phys. (N.Y.)} {\twelvebf 168} (1986) 148;
H.-Th.Elze and U.Heinz, {\twelveit Phys.Rep.} {\twelvebf 183} (1989) 81.
\itemitem{26.}  E.Braaten, in {\twelveit Hot Summer Daze},
A.Gocksch and R.Pisarski (eds.)
(World Scientific, Singapore, 1992).
\itemitem{27.} R.I.Nepomechie, {\twelveit Phys.Rev.} {\twelvebf D33}
(1986) 3670; D.Karabali, Q-H.Park, \break
H.J.Schnitzer and Z.Yang, {\twelveit
Phys.Lett.} {\twelvebf 216B} (1989) 307; D.Karabali and H.J.Schnitzer,
{\twelveit Nucl.Phys.} {\twelvebf B329} (1990) 649; K.Gawedzki and
A.Kupianen, {\twelveit Phys.Lett.} {\twelvebf 215B} (1988) 119;
{\twelveit Nucl.Phys.} {\twelvebf B320} (1989) 649.
\itemitem{28.} R.Pisarski, {\twelveit Physica} {\twelvebf A158}
(1989) 146; T.Appelquist and
R.Pisarski, {\twelveit Phys. Rev} {\twelvebf D23} (1981) 2305.
\itemitem{29.}  J.P.Blaizot and E.Iancu, {\twelveit Phys.Lett.}
{\twelvebf B326} (1994) 138.
\itemitem{30.} V.P.Nair, Preprint CCNY-HEP 94/4, May 1994.
\itemitem{31.} A.D. Linde, {\twelveit Phys.Lett.} {\twelvebf B96}
(1980) 289; D. Gross, R. Pisarski and L. Yaffe,
{\twelveit Rev.Mod.Phys.} {\twelvebf 53} (1981) 43.
\itemitem{32.} Y.Nambu and G.Jona-Lasinio, {\twelveit Phys.Rev.}
{\twelvebf 122} (1961) 345; {\twelveit ibid.} {\twelvebf 124} (1961)
246.

\bye